\documentclass[twocolumn,preprintnumbers,amsmath,amssymb,superscriptaddress]{revtex4-1}
\usepackage{ulem}
\usepackage[dvips]{graphicx}
\usepackage{color}
\usepackage{bm}
\usepackage{subcaption}
\begin{document}
\title{Constrained correlated-Gaussians for hyperspherical calculations} 
\author{Y. Suzuki}
\affiliation{Department of Physics, Niigata University, Niigata 950-2181, Japan}
\affiliation{RIKEN Nishina Center, Wako 351-0198, Japan}
\author{K. Varga}
\affiliation{Department of Physics and Astronomy, Vanderbilt University, Nashville, Tennessee 37235, USA}
\date{\today}
\begin{abstract}
We formulate a hyperspherical approach within standard configuration 
interaction calculations aiming at a description of 
large-scale dynamics of $N$-particle system.    
The channel wave function and the adiabatic channel energy are determined by 
solving a hyperradius-constrained eigenvalue problem of the adiabatic 
Hamiltonian.  
The needed matrix elements are analytically evaluated 
using correlated Gaussians with good orbital angular momentum and 
parity. The 
feasibility of the approach is tested in three-$\alpha$ system. 
A spectrum of the adiabatic channel energies is determined depending on 
the degree of localization of the basis functions.
\end{abstract}
\maketitle

\section{Introduction}
\label{intro}

The hyperspherical coordinate system is a natural extension of the 
three-dimensional spherical polar coordinates to a set of 
$N$-particle coordinates. The hyperspherical approach  
attempts to solve an $N$-particle  
Schr\"odinger equation by expressing the total wave function as a product 
of the hyperradial and hyperangular parts and can be used to
solve bound and continuum state problems. 

The main advantage of the hyperspherical method is that it provides
a unified framework to describe quantum dynamics of complex reactions
such as decay, fusion or fission. In other methods the choice of
relevant coordinates is not trivial. For example, in nuclear fusion initially
the relative distance between the nuclei might be the most important
coordinate, but later other coordinates will be more suitable and
necessary. In the hyperspherical approach
the hyperradius captures all features of the complicated dynamical
processes and describe dynamical properties of the system emerging at
different hyperradial distances. 

Two realizations of the hyperspherical approach are widely used (see 
Refs.~\cite{zhukov93,lin95,krivec98,nielsen01,greene17} for reviews on the 
hyperspherical approach and its applications). 
In one approach the hyperangular part is expanded 
in terms of the hyperspherical harmonics that are eigenfunctions of the 
hyperangular kinetic-energy operator, and a coupled  
hyperradial equation is solved by including the interaction of the particles. 
In another approach, 
often called the adiabatic hyperspherical approach, the adiabatic Hamiltonian 
consisting of the hyperangular 
kinetic energy and the interaction potential is diagonalized first to obtain 
the adiabatic channel energies and channel wave functions. The total wave 
function is then expanded in terms of the basis set of the adiabatic channel 
wave functions. The adiabatic channel energies 
give hints on how the system responds as a function of hyperradial distances.

The advantage of the first approach is that  the hyperspherical 
harmonics are known, but two difficulties may hinder the 
application. One problem is that convergence of the 
hyperspherical harmonics expansion is slow  
even for short-range potentials~\cite{thompson00}, and it becomes 
prohibitively slow when a long-range potential, like the 
Coulomb coupling potential, acts at large hyperradial distances. 
This slowness is related to the fact that the hyperangular 
kinetic energy and the interaction potential do not 
commute~\cite{macek68,kvitsinsky91}. 
The slow convergence causes huge discrepancies, e.g. 
in the triple-$\alpha$ reaction rate at low 
temperatures~\cite{nguyen13,ishikawa13,suno16,suno15}. 
Another problem is that solving the coupled differential 
equation in the hyperradial 
coordinate may become hard when a number of avoided crossings occur in 
the potential energy curves. 

Although the basic idea of the hyperpsherical method is not limited 
to three-body systems, its extension to 
more-particle system is impeded
by the lack of appropriate basis functions that can be flexibly used in the hyperspherical 
calculation. References~\cite{barnea00,barnea03,barnea10,bacca12,timofeyuk02,timofeyuk08,gattobigio11} discuss
recent developments in going beyond the three-body problems. 

In both realization of the hyperspherical approach, 
one calculates the matrix element of an operator ${\cal O}$, 
\begin{align}
\langle \Psi' |{\cal O}|\Psi \rangle_{\rho=R},
\label{me.angle}
\end{align}
where $\langle \ldots \rangle_{\rho=R}$ indicates that the matrix element 
is to be evaluated by integrating in all the coordinates but the hyperradius 
$\rho$, which is fixed to $R$. The integral of type~(\ref{me.angle}) is 
hard to evaluate in general because specifying the hyperangle coordinates 
for the $N$-particle system 
is considerably involved and integrating in those coordinates requires 
many-dimensional integrations. Although some progress has recently been made 
with correlated Gaussian (CG) basis 
functions~\cite{stecher09,rittenhouse11,rakshit12,daily14}, 
the total orbital angular momentum is limited to $L=0$ and 1.

In this paper we will examine the possibility of using the 
CG as suitable basis functions in hyperspherical
calculations.  The CG proposed many years ago~\cite{boys60,singer60} 
is extended to describe motion with non-zero total orbital angular 
momentum, especially with the help of the  global 
vector representation~\cite{varga95,book,suzuki98}. 
Together with the stochastic variational method~\cite{varga94,varga95,book} 
to select efficiently the parameters of the CG, many problems 
have accurately been solved with the CG. See, e.g. 
Refs.~\cite{mitroy13,horiuchi14,mikami14} for some recent applications of CG. 

We attempt to formulate the hyperspherical approach 
in standard configuration interaction calculations. 
Following the spirit of the second realization of the hyperspherical
approach, we set up a number of basis functions 
that are expected to be important at $\rho\approx R$, calculate the 
matrix elements of the adiabatic Hamiltonian using the  
full coordinate integration instead of Eq.~(\ref{me.angle}), 
and determine both the 
channel wave function and the adiabatic channel energy at $\rho\approx R$ 
by solving a hyperradius-constrained eigenvalue equation. We show that 
this scheme can be achieved using the CG basis functions. 
The emphasis of this paper is not on solving a specific problem 
with the hyperspherical approach but on carefully examining its 
feasibility and discussing problems that may occur.    

We present our formulation in Sec.~\ref{framework}, and show  
in Sec.~\ref{general.case} how to evaluate the needed matrix elements. 
In Sec.~\ref{3alpha.example} we test  our method in three-$\alpha$ system 
that is the simplest possible system but 
contains all the complexities nevertheless. Section~\ref{summary} is 
a summary and discussions.

\section{Schr\"odinger equation in hyperspherical approach}
\label{framework}

\subsection{Hyperspherical coordinates}

Let $\bm r_i\, (i=1,\ldots,N)$ denote the position coordinate of the $i$th particle. The mass $m$ of all particles is assumed to be the same, 
although the case of unequal mass 
can be treated by defining mass-scaled coordinates. 
We define  a set of relative coordinates, $\bm x_i \ (i=1,\ldots, N-1)$, 
\begin{align}
\bm x_i=\sqrt{\frac{i}{i+1}}\Big(\bm r_{i+1}-\frac{1}{i}\sum_{j=1}^{i}\bm r_j\Big).
\end{align}
The set $\bm x_i$ together with the center of mass (c.m.) coordinate, $\bm x_N \equiv {\bm R}_{\rm cm}=\sum_{i=1}^N{\bm r}_i/N$,  
defines a transformation matrix $U$ from the single-particle 
coordinates to the 
relative and c.m. coordinates: 
\begin{align}
\bm x_i=\sum_{j=1}^N U_{ij}\bm r_j\ \  (i=1,\ldots, N).
\end{align}
Conversely, $\bm r_i$ is expressed as 
$\bm r_i=\sum_{j=1}^N {U^{-1}}_{ij}\bm x_j$.  

The square of the hyperradius $\rho$ is defined by 
\begin{align}
\rho^2&=\sum_{i=1}^N(\bm r_i-\bm R_{\rm cm})^2
=\frac{1}{N}\sum_{j>i=1}^N(\bm r_i-\bm r_j)^2,
\label{def.rho1}
\end{align}
which is equal to 
\begin{align}
\rho^2=\sum_{i=1}^{N-1}\bm x_i^2.
\label{def.rho}
\end{align}
Let $\Omega$ denote a set of the hyperangle coordinates constructed from 
dimensionless coordinates, $\bm \xi_i=\bm x_i/\rho \ (i=1,\ldots, N-1)$. They  
are constrained as $\sum_{i=1}^{N-1}\bm \xi_i^2=1$. 
The volume element for integration excluding ${\bm R}_{\rm cm}$ is 
\begin{align}
d\bm x\equiv d\bm x_1 \ldots d\bm x_{N-1}=\rho^{d-1}d\rho d\Omega, 
\label{vol.element}
\end{align}
where 
\begin{align}
d=3(N-1)
\end{align}
is the degree of freedom excluding the c.m. motion.

Since $\rho^2/N$ is the mean-square-radius operator, $\rho$ 
measures the global size of the system. Or 
$\rho$ is a kind of collective coordinate 
responding to a large-scale change of 
the system~\cite{suzuki15}.  
Suppose that the system develops into $f$ subsystems or clusters, each of 
which consists of $N_i$ particles ($\sum_{i=1}^f N_i=N$). $\rho^2$ is  
rewritten as 
\begin{align}
\rho^2=\sum_{i=1}^f\rho_{i}^2+\rho_{\rm rel}^2
\end{align}
with
\begin{align}
\rho_i^2=\sum_{j=1}^{N_i}({\bm r}_{(i-1)+j}-{\bm R}_i)^2,\ \ \ 
\rho_{\rm rel}^2=\sum_{i=1}^f N_i({\bm R}_i-{\bm R}_{\rm cm})^2,
\end{align}
where $(i-1)=\sum_{k=1}^{i-1}N_k$ with $(0)=0$, and 
${\bm R}_i$ is the c.m. coordinate of the $i$th cluster. $\rho_i$ is 
the hyperradius of the $i$th cluster, and $\rho_{\rm rel}$ stands for 
the hyperradius 
that measures the spatial extension of the relative motion of the clusters. 
In such phenomena that include the formation of $f$ subsystems, the 
contribution of $\sum_{i=1}^f\rho_{i}^2$ to $\rho^2$ remains finite, whereas 
$\rho^2_{\rm rel}$ takes an increasingly large value as $\rho$ increases.  
Moreover, since $\rho^2$ is invariant with respect to the number of 
clusters $f$, 
the hyperspherical coordinates have the unique advantage that they can 
treat any decomposition of the system in a unified way.

\subsection{Equation of motion in hyperspherical approach}
\label{eq.motion}

The Hamiltonian $H$ of the system consists of the kinetic energy 
$T$ and the interaction potential $V$:
\begin{align}
H=T+V.
\end{align}
With the c.m. kinetic energy being subtracted, $T$ reads 
\begin{align}
T=-\frac{\hbar^2}{2m}\sum_{i=1}^{N-1}\frac{\partial^2}{\partial \bm x_i^2}, 
\end{align}
and it is separated into hyperradial ($T_{\rho}$) 
and hyperangular ($T_{\Omega}$) parts, $T=T_{\rho}+T_{\Omega}$:
\begin{align}
T_{\rho}=-\frac{\hbar^2}{2m}\Big(
\frac{\partial^2}{\partial \rho^2}+\frac{d-1}{\rho}\frac{\partial}{\partial \rho}\Big),\ \ \ 
T_{\Omega}=\frac{\hbar^2 \Lambda^2} {2m \rho^2},
\label{decomp.kine}
\end{align}
where $\Lambda^2$ is the squared grand angular momentum operator that can in principle 
be expressed in terms of the hyperangle coordinates and their derivatives. 
The adiabatic Hamiltonian
\begin{align}
H_{\Omega}=T_{\Omega}+V
\end{align}
is a kind of a generalized potential. As usual, 
$V$ and consequently $H_{\Omega}$ is assumed to contain no derivative 
operator with respect to $\rho$. 

Let the total wave function $\Psi^{JM\pi}$ of the system be labeled by  
the total angular momentum $J$, its $z$ component $M$, the parity $\pi$.   
The Schr\"odinger equation, $H\Psi^{JM\pi}=E^{J\pi}\Psi^{JM\pi}$, reads as
\begin{align}
(T_{\rho}+H_{\Omega})\Psi^{JM\pi}=E^{J\pi}\Psi^{JM\pi}.
\label{sch.eq}
\end{align}
The channel wave function $\Phi_{\nu}^{JM\pi}(\rho, \Omega)$ and the adiabatic 
channel energy (or adiabatic potential) $U_{\nu}^{J\pi}(\rho)$ are defined 
by solving the eigenvalue problem of $H_{\Omega}$, 
\begin{align}
H_{\Omega}\Phi_{\nu}^{JM\pi}(\rho, \Omega)=U_{\nu}^{J\pi}(\rho)\Phi_{\nu}^{JM\pi}(\rho, \Omega).
\label{adiabatic.eq}
\end{align}
In order to exploit the fact that $\rho$ can be treated as a $c$-number in 
Eq.~(\ref{adiabatic.eq}), we need the matrix element 
$\langle \Phi_{\nu}^{JM\pi}|H_{\Omega}|\Phi_{\nu'}^{JM\pi}\rangle_{\rho=R}$. 
Since its evaluation is, however, hard as already mentioned, we take a 
different route. 

We set up a number of independent basis functions 
$\Phi^{JM\pi}_{l}\ (l=1, \ldots, M)$ that satisfy 
\begin{align}
\langle \Phi^{JM\pi}_{l}|\rho^2|\Phi^{JM\pi}_{l}\rangle=R^{2},
\label{rho2.constraint}
\end{align}
and assume that the $\nu$th `channel wave function' $\Phi^{JM\pi}_{R \nu}$ 
at $\langle \rho ^2\rangle= R^2$ is given as a combination of the basis 
functions  
\begin{align}
\Phi^{JM\pi}_{R \nu}=\sum_{l=1}^{M}c^{J\pi}_{R \nu,l}\Phi^{JM\pi}_{l}.
\label{channel.wf} 
\end{align}
The coefficients $c^{J\pi}_{R \nu,l}$  
are determined by solving the constrained eigenvalue problem 
\begin{align}
\langle \Phi^{JM\pi}_{l} | H_{\Omega}  - U_{\, R\nu}^{J\pi} |\Phi^{JM\pi}_{R\nu} \rangle=0 \ \ \ (l=1,\ldots, M), 
\label{eig.channel}
\end{align}
subject to 
\begin{align}
\langle \Phi^{JM\pi}_{R\nu} |\rho^2| \Phi^{JM\pi}_{R\nu} \rangle=R^2.
\label{rho2.constr}
\end{align}
Both $\Phi^{JM\pi}_{l}$ and 
$\Phi^{JM\pi}_{R \nu}$ are normalized, and 
the matrix elements in Eqs.~(\ref{rho2.constraint}), (\ref{eig.channel}), 
and (\ref{rho2.constr}) 
are evaluated by integrating in all the coordinates.   
Appendix~\ref{eigenvalue.constraint} shows 
how to determine the adiabatic channel energies 
$U^{J\pi}_{R\nu}$ at $R$ and mutually orthogonal channel wave functions 
$\Phi^{JM\pi}_{R\nu}$.

We calculate the channel wave functions at a number of mesh points $R_i$ and 
assume the total wave function to be approximated by their combinations 
\begin{align}
\Psi^{JM\pi}=\sum_{i \nu}\chi^{J\pi}_{i \nu}\Phi^{JM\pi}_{R_i \nu}.
\end{align}
Equation~(\ref{sch.eq}) reduces to the following 
equation for $\chi^{J\pi}_{i \nu}$:
\begin{align}
&\sum_{j\nu'}\langle \Phi^{JM\pi}_{R_i \nu}|T_{\rho}+ H_{\Omega}-E^{J\pi}|\Phi^{JM\pi}_{R_j \nu'}\rangle  \chi^{J\pi}_{j \nu'}=0 \notag \\
&\qquad \qquad \qquad \qquad  \qquad \qquad {\rm for\ all}\ i\  {\rm and}\ \nu.
\label{eq.to.be.solved}
\end{align}

The condition~(\ref{rho2.constraint}) is necessary  
to look for a suitable basis set 
at $ R$ because each piece of $H_{\Omega}$ shows  
different $\rho$-dependence and hence such set may change depending on 
$\rho$. Short-ranged interactions in $V$ become important 
in the region of small $\rho$, while long-ranged interactions like the Coulomb 
potential contribute at large $\rho$ as well. $T_{\Omega}$ is also long-ranged. 
Moreover, since the Coulomb potential and $T_{\Omega}$ do not 
commute each other, one has to take account of 
both terms simultaneously~\cite{macek68}. 

As a measure of the localization of a wave function 
$\Phi$, we introduce the standard deviation $\sigma$ of $\rho^2$:  
\begin{align}
\sigma^2=\frac{\langle \Phi |(\rho^2-\langle \Phi |\rho^2|\Phi \rangle)^2|\Phi \rangle}{\langle \Phi |\rho^2|\Phi \rangle^2}=
\frac{\langle \Phi |\rho^4|\Phi \rangle}{\langle \Phi |\rho^2|\Phi \rangle^2}-1.
\label{stand.dev}
\end{align}
$\Phi^{JM\pi}_{R\nu}$ is obtained as a combination of $\Phi^{JM\pi}_{l}$s. Therefore, 
even though $\Phi^{JM\pi}_{l}$s are all set to have $\sigma$'s within a certain range, 
it may happen that the $\sigma$ value of $\Phi^{JM\pi}_{R\nu}$ is far beyond its range. 
To obtain $U_{\, R\nu}^{J\pi}$ around $\rho \approx R$, it is useful to check 
the $\sigma$ value of $\Phi^{JM\pi}_{R\nu}$. We will discuss this problem later. 

Once $\Phi^{JM\pi}_{R_i\nu}$s are determined at $R_i$'s, Eq.~(\ref{eq.to.be.solved}) 
can be solved in a standard linear algebra. Note that the matrix 
element of $T_{\rho}$ is already available at the stage of solving the 
eigenvalue problem of $H_{\Omega}$. This is in sharp contrast to the 
standard hyperspherical method where no hyperradial function is employed and 
thus one has to use numerical differentiations with respect to $\rho$ or e.g. 
slow variable discretization method~\cite{tolstikhin96,suno11}.  

In what follows we omit the superscripts $JM\pi$.

\section{Correlated Gaussian as hyperspherical basis function}
\label{general.case}

\subsection{Correlated Gaussian and its generating function}

We adopt the CG as the basis function. 
We use matrix notations to make equations compact. 
For example, $\bm x$ denotes a column vector 
of dimension $(N-1)$ whose $i$th element is $\bm x_i$. A tilde  
symbol $\widetilde{\ \ \ }$ indicates 
a transpose of a column vector or a matrix, e.g. $\widetilde{\bm x}$ is 
the row 
vector and $\rho^2$ may be written as $\widetilde{\bm x}\bm x$, where 
the scalar product of 3-dimensional vectors is implicitly understood: 
$\widetilde{\bm x}\bm x=\sum_{i=1}^{N-1}{\bm x}_i\cdot {\bm x}_i
=\sum_{i=1}^{N-1}{\bm x}_i^2$. 

The CG with the total orbital angular momentum $L$ and its 
$z$ component $M$ reads 
\begin{align}
f^{uA}_{KLM}(\bm x)={\cal N}^{uA}_{KL}|\widetilde{u}\bm x|^{2K+L}Y_{LM}(\widehat{\widetilde{u}\bm x})e^{-\frac{1}{2}\widetilde{\bm x}A\bm x},
\label{def.CG}
\end{align}
where a column vector $u=(u_i)$ of  dimension $(N-1)$  
and a symmetric, positive-definite $(N-1)\times (N-1)$ 
matrix $A=(A_{ij})$ are both (variational) parameters to characterize the CG.  
Both $A$ and $u$ are assumed to be real in this paper. 
The exponential part, $e^{-\frac{1}{2}\widetilde{\bm x}A\bm x}$, is invariant under the 
coordinate rotation, whereas the spherical harmonics $Y_{LM}$ describes the rotational 
motion through the global vector, 
$\widetilde{u}\bm x=\sum_{i=1}^{N-1}u_i{\bm x}_i$~\cite{varga95,book,suzuki98,suzuki08,aoyama12}. $\widehat{\widetilde{u}\bm x}$ stands for the polar and azimuthal angles of $\widetilde{u}\bm x$. ${\cal N}^{uA}_{KL}$ is the normalization constant determined from  
$\langle f^{uA}_{KLM}|f^{uA}_{KLM}\rangle=1$. 

$K$ is a non-negative integer parameter related to the localization 
in $\rho$ motion of the CG~\cite{suzuki17}. 
It should be noted that the CG has simple hyperadial dependence 
\begin{align}
f^{uA}_{KLM}(\bm x) \sim \rho^{\kappa}e^{-\frac{1}{2} \rho^2 \widetilde{\bm \xi}A\bm \xi} \ \ \ (\kappa=2K+L),
\label{cg.rho-dep}
\end{align}
where $\bm \xi=({\bm \xi}_i)$ is a column vector of 
dimension $(N-1)$. This simplicity makes it easy to calculate 
the matrix element of $T_{\rho}$.  

Let us introduce the generating function for the CG, 
\begin{align}
g({\bm s}, A, {\bm x})=e^{-\frac{1}{2}\widetilde{\bm x}A\bm x+ \widetilde{\bm s}\bm x},  
\end{align}
where $\bm s=({\bm s}_i)$ is a column vector of dimension $(N-1)$ consisting of 
3-dimensional vector ${\bm s}_i$.  With a choice of ${\bm s}_i=\alpha u_i {\bm e}$, 
where $\alpha$ is an auxiliary real parameter and $\bm e$ is a 
three-dimensional 
unit vector ($\bm e^2=\bm e \cdot \bm e=1$), 
the CG is generated as follows\cite{varga95,book}: 
\begin{align}
f^{uA}_{KLM}(\bm x)&=\frac{{\cal N}^{uA}_{KL}}{B_{KL}}\int d{\bm e} Y_{LM}(\hat{{\bm e}})\notag \\
&\times \Big(\frac{d^{2K+L}}{d\alpha^{2K+L}}g(\alpha u\bm e, A,\bm x)\Big)_{\alpha=0}
\label{cg.gfn}
\end{align}
with 
\begin{align}
B_{KL}=\frac{4\pi (2K+L)!}{2^KK!\, (2K+2L+1)!!}.
\label{def.BKL}
\end{align}
Here $(\ \ \ )_{\alpha=0}$ indicates that $\alpha$ 
is set to zero after the differentiation.

\subsection{Basic matrix elements}
\label{CGme.vs.GHF}

The CG matrix elements for various operators are available in the 
literature~\cite{varga95,book,suzuki98,suzuki17}. We 
recapitulates the basic procedure to derive them  
with emphasis on the relationship to the 
Gauss hypergeometric function (GHF)~\cite{bateman53,abramowitz70}, which  
has hitherto never been recognized. 

Applying Eq.~(\ref{cg.gfn}) leads to the CG matrix element: 
\begin{align}
&\langle f^{u'A'}_{K'LM} | {\cal \hat O} |f^{uA}_{KLM} \rangle \notag \\
&=\frac{{\cal N}^{u'A'}_{K'L}}{B_{K'L}}\frac{{\cal N}^{uA}_{KL}}{B_{KL}}
\int d{\bm e'} \int d{\bm e}\, Y_{LM}^*(\hat{{\bm e}'})Y_{LM}(\hat{{\bm e}})\notag \\
& \times \Big(\frac{d^{2K'+L+2K+L}}{d\alpha'^{2K'+L}d\alpha^{2K+L}}
\int d\bm x \, e^{-\widetilde{\bm x}B\bm x+\widetilde{\bm v}\bm x }{\cal O}(\bm x)
\Big)_{\substack{\alpha=0 \\ \alpha'=0}}.
\label{CG.me.general.formula}
\end{align}
Here ${\cal O}(\bm x)$ is determined by acting ${\cal \hat O}$ on 
$g(\bm s; A,\bm x)$ or $f^{uA}_{KLM}(\bm x)$. The matrix $B$ and the 
vector $\bm v$ are defined by 
\begin{align}
B=\frac{1}{2}(A+A'),\ \ \ \ \ \bm v= \bm s  + {\bm s'}, 
\end{align}
where $\bm s=\alpha u \bm e$ and ${\bm s'}=\alpha' u' {\bm e'}$.

For a class of operators, the integral in Eq.~(\ref{CG.me.general.formula}) over the whole region of $\bm x$  takes the form
\begin{align}
\int d\bm x \, e^{-\widetilde{\bm x}B\bm x+\widetilde{\bm v}\bm x }{\cal O}(\bm x)&={\cal P_O}\Big(\frac{\pi^{N-1}}{{\rm det}B}\Big)^{\frac{3}{2}}e^{\frac{1}{4}\widetilde{\bm v}B^{-1}\bm v}.
\label{basic.integral}
\end{align}
Appendix~\ref{CG.m.e} lists some examples of 
${\cal O}(\bm x)$ and ${\cal P_O}$. 
In all those cases, ${\cal {P_O}}$ consists of terms with the form    
\begin{align}
T_{kk'l}(u'A',uA){\alpha}^{2k}  {\alpha'}^{2k'} (\alpha \alpha' \bm e\cdot \bm e')^l,
\label{PO.general}
\end{align}
each of which is characterized by non-negative integers, $k$, $k'$, $l$, 
and the coefficient $T_{kk'l}(u'A',uA)$. 
The exponent in Eq.~(\ref{basic.integral}) is 
\begin{align}  
\frac{1}{4}\widetilde{\bm v}B^{-1}\bm v=p\alpha^2+p'\alpha'^2+q\alpha \alpha' 
\bm e\cdot {\bm e}',
\end{align}
where
\begin{align}
p=\frac{1}{4}\widetilde{u}B^{-1}u,\ \ \ p'=\frac{1}{4}\widetilde{u'}B^{-1}u',\ \ \ q=\frac{1}{2}\widetilde{u}B^{-1}u'.
\label{def.pp'q}
\end{align}
$\widetilde{u}Au'$ or $(\widetilde{u}Au')$ stands for the inner product, 
$\sum_{i,j=1}^{N-1}u_iA_{ij}u'_j$.   
Expanding $e^{\frac{1}{4}\widetilde{\bm v}B^{-1}\bm v}$ in a power series of 
$\alpha$ and $\alpha'$, and combining it with the term of 
Eq.~(\ref{PO.general}), we perform the 
operation in Eq.~(\ref{CG.me.general.formula}), obtaining the contribution 
of term~(\ref{PO.general}) to the matrix element as follows: 
\begin{align}
&\langle f^{u'A'}_{K'LM} | {\cal O} | f^{uA}_{KLM} \rangle 
\sim  \frac{{\cal N}^{u'A'}_{K'L}}{B_{K'L}}\frac{{\cal N}^{uA}_{KL}}{B_{KL}}\Big(\frac{\pi^{N-1}}{{\rm det}B}\Big)^{\frac{3}{2}}
\notag \\ 
& \times T_{kk'l}(u'A',uA) (2K+L)!(2K'+L)! \notag \\
&\times 
\sum_{n=n_0}^{n_1} \frac{p^{K-k-n}{p'}^{K'-k'-n}q^{2n+L-l}B_{nL}}{(K-k-n)!(K'-k'-n)!(2n+L-l)!},
\label{cgme.general}
\end{align}
where $n_1$ and $n_0$ are given by
\begin{align}
&n_1={\rm min}(K-k,K'-k'),\notag \\
&n_0=\Big\{
\begin{array}{cc}
0 &  {\rm for \ }L \geqq l \\
\big[\frac{l-L+1}{2}\big] & {\rm for \ }l > L. 
\end{array}
\end{align}
Here Gauss's symbol $[x]$ stands for the greatest integer that is less than or equal to $x$. 

The sum in Eq.~(\ref{cgme.general}) can be expressed with the GHF as follows. 
By using $B_{nL}$~(\ref{def.BKL}),  the sum reduces to    
\begin{align}
&\sum_{n=n_0}^{n_1} \frac{p^{K-k-n}{p'}^{K'-k'-n}q^{2n+L-l}B_{nL}}{(K-k-n)!(K'-k'-n)!(2n+L-l)!}\notag \\
&= \frac{4\pi \, p^{K-k}{p'}^{K'-k'}q^{L-l}(2z)^{n_0}}{(K-k-n_0)!(K'-k'-n_0)!(2L+2n_0+1)!!}\notag \\
&\times
 \sum_{m=0}^{n_1-n_0}\frac{(-K+k+n_0)_m (-K'+k'+n_0)_m}{m!(L+n_0+\frac{3}{2})_m}P^{L,n_0}_l(m)z^m,
\label{summation.formula.step}
\end{align}
where $(a)_m$ is Pochhammer's symbol
\begin{align}
(a)_m=\frac{\Gamma(a+m)}{\Gamma(a)}
\label{poch}
\end{align}
expressed with the Gamma function $\Gamma$.
If $a$ is negative, $(a)_m=(-1)^m \Gamma(-a+1)/\Gamma(-a-m+1)$. If $a$ is 
a negative integer, $a=-k$,  $(-k)_m=0$ for $m > k$.
$z$ in Eq.~(\ref{summation.formula.step}) is defined by 
\begin{align}
z=\frac{q^2}{4pp'}=\frac{(\widetilde{u}B^{-1}u')^2}{(\widetilde{u}B^{-1}u)(\widetilde{u'}B^{-1}u')},
\end{align}
and takes a value in the interval $[0,1]$.  
$P^{L,n_0}_l(m)$ in Eq.~(\ref{summation.formula.step}) is 
a polynomial of $m$ with the order $l-n_0$,
\begin{align}
P^{L,n_0}_l(m)=\frac{m!(2m+L+2n_0)!}{(m+n_0)!(2m+L-l+2n_0)!}.
\label{dif.op.P}
\end{align}
Because of $m^iz^m=(z\frac{d}{dz})^iz^m$ for any non-negative 
integer $i$, $P^{L,n_0}_l(m)z^m$ may be replaced by $P^{L,n_0}_l\big(z\frac{d}{dz}\big)z^m$, which 
makes Eq.~(\ref{summation.formula.step}) further compact: 
\begin{align}
&\sum_{n=n_0}^{n_1} \frac{p^{K-k-n}{p'}^{K'-k'-n}q^{2n+L-l}B_{nL}}{(K-k-n)!(K'-k'-n)!(2n+L-l)!}\notag \\
&\ \ =\frac{4\pi \, p^{K-k}{p'}^{K'-k'}q^{L-l}(2z)^{n_0}}{(K-k-n_0)!(K'-k'-n_0)!(2L+2n_0+1)!!}\notag \\
&\ \ \times P^{L,n_0}_l\big(z\frac{d}{dz}\big)\gamma_{K-k-n_0, K'-k'-n_0, L+n_0}(z).
\label{summation.formula}
\end{align}
Here $\gamma_{K,K',L}(z)$, introduced in Ref.~\cite{suzuki17}, 
is nothing but the GHF 
\begin{align}
\gamma_{K,K',L}(z)={}_2F_{1}(-K,-K';L+\textstyle{\frac{3}{2}};z),
\end{align}
which is actually a polynomial of $z$ with the order $\min(K,K')$ because   
$K$ and $K'$ are both non-negative integers in the present case. 

Equations~(\ref{cgme.general}) and (\ref{summation.formula}) constitute 
a basic formula to calculate the matrix element.  Let us 
consider the overlap matrix element, for which ${\cal O}(\bm x)=1$, 
${\cal P_{O}}=1$, $k=k'=l=0$, $T_{000}(u'A',uA)=1$, leading to 
\begin{align}
&\langle f^{u'A'}_{K'LM} | f^{uA}_{KLM} \rangle \notag \\
&=\frac{{\cal N}^{u'A'}_{K'L}}{B_{K'L}}\frac{{\cal N}^{uA}_{KL}}{B_{KL}}
\Big(\frac{\pi^{N-1}}{{\rm det}B}\Big)^{\frac{3}{2}} (2K+L)!(2K'+L)! \notag \\
&\times \frac{4\pi \, p^K p'^{K'} q^L}{K! K'! (2L+1)!!} \gamma_{K, K', L}(z).
\label{fOf}
\end{align}
In the diagonal case of $u'=u, \, A'=A,\, K'=K$, $z$ is unity and 
$\gamma_{K, K, L}(1)$ is easily obtained by using 
\begin{align}
 {}_2F_{1}(a,b;c;1)&=\frac{\Gamma(c)\Gamma(c-a-b)}{\Gamma(c-a)\Gamma(c-b)},
\label{GHFz=1}
\end{align}
which is valid provided that ${\rm Re}\, (c-a-b)>0$. 
The normalization constant is then given by 
\begin{align}
{\cal N}^{uA}_{KL}
&=\sqrt{\frac{2({\rm det}A)^{\frac{3}{2}}} 
{\sqrt{\pi}^{3(N-2)} \Gamma(2K+L+\frac{3}{2}) (\widetilde{u}{A}^{-1}u)^{2K+L}}}.
\label{normalization.const}
\end{align}
Substitution of 
Eqs.~(\ref{def.BKL}), (\ref{def.pp'q}), and (\ref{normalization.const}) 
into Eq.~(\ref{fOf}) and the use of Eq.~(\ref{GHFz=1}) 
completes the overlap matrix element:
\begin{align}
&\langle f^{u'A'}_{K'LM} | f^{uA}_{KLM} \rangle \notag \\
&=\Big(\frac{{\rm det}AA'}{({\rm det} B)^2}\Big)^{\frac{3}{4}}\Big(\frac{\widetilde{u}B^{-1}u}{\widetilde{u}A^{-1}u}\Big)^{\frac{1}{2}(2K+L)}
\Big(\frac{\widetilde{u'}B^{-1}u'}{\widetilde{u'}A'^{-1}u'}\Big)^{\frac{1}{2}(2K'+L)}
\notag \\
&\  \times \Big(\frac{\widetilde{u}{B}^{-1}u'}{|\widetilde{u}{B}^{-1}u'|}\sqrt{z}\Big)^L 
\frac{\gamma_{K, K', L}(z)}{\sqrt{\gamma_{K,K,L}(1)\gamma_{K',K',L}(1)}}. 
\label{ff.overlap}
\end{align}

Combining Eqs.~(\ref{cgme.general}), (\ref{summation.formula}), and 
(\ref{fOf}) enables us to express the 
contribution of term~(\ref{PO.general}) in relation to 
the overlap matrix element: 
\begin{align}
&\langle f^{u'A'}_{K'LM} | {\cal O} | f^{uA}_{KLM} \rangle 
\sim \langle f^{u'A'}_{K'LM} | f^{uA}_{KLM} \rangle \notag \\
&\ \ \ \times   T_{kk'l}(u'A',uA) p^{-k}p'^{-k'}q^{-l}F^{KK'L}_{kk'l}(z),
\label{me.vs.ovlme}
\end{align}
where
\begin{align}
&F^{KK'L}_{kk'l}(z)
=\frac{K!\, K'!}{(K-k-n_0)! \,(K'-k'-n_0)! \, (L+\frac{3}{2})_{n_0}} \notag \\
&\ \times \frac{z^{n_0}}{\gamma_{K, K', L}(z)}P^{L,n_0}_l\big(z\frac{d}{dz}\big)
\gamma_{K-k-n_0, K'-k'-n_0, L+n_0}(z).
\label{function.FKKL}
\end{align}
Equations~(\ref{ff.overlap}), (\ref{me.vs.ovlme}), and (\ref{function.FKKL}) 
give a powerful formula for the matrix element. We only need 
to determine $T_{kk'l}(u'A',uA)$, which contributes to the matrix element 
provided that both $K-k-n_0$ and $K'-k'-n_0$ are non-negative. 

\begin{table}
\caption{$F^{KKL}_{kk'l}(1)$ for some sets of 
$(k,k',l)$. See Eq.~(\ref{function.FKKL}). Note that 
$F^{KKL}_{k'kl}(1)=F^{KKL}_{kk'l}(1)$. $M_1$ and $M_2$ stand for 
$M_1=K+L+\frac{1}{2}$ and $M_2=2K+L+\frac{1}{2}$, respectively. 
 } 
\label{value.F.z=1}
\begin{tabular}{ccccccl}
\hline\hline
$k$& $k'$ & $l$ &&&& $F^{KKL}_{kk'l}(1)$ \\
\hline
\vspace{1mm}
0 & 0 & 0 &&&& 1\\
\vspace{1mm}
1 & 0 & 0 &&&& $K\frac{ M_1}{M_2}$ \\
\vspace{1mm}
0 & 0 & 1 &&&& $2\frac{{M_1}^{\!2}}{M_2}-L-1$ \\
\vspace{1mm}
2 & 0 & 0 &&&& $K(K-1)\frac{M_1(M_1-1)}{M_2(M_2-1)}$ \\
\vspace{1mm}
1 & 1 & 0 &&&& $K^2\frac{{M_1}^{\!2}}{M_2(M_2-1)}$\\
\vspace{1mm}
1 & 0 & 1 &&&& $2K\frac{{M_1}^{\!2}(M_1-1)}{M_2(M_2-1)}-K(L+1)\frac{M_1}{M_2}$\\
\vspace{1mm}
0 & 0 & 2 &&&& $4\frac{{M_1}^{\!2}(M_1-1)^2}{M_2(M_2-1)}-2(2L+1)\frac{{M_1}^{\!2}}{M_2}+(L+1)(L+2)$\\
\hline\hline
\end{tabular}
\end{table}

Small values of $k$, $k'$, and $l$ are usually needed. 
For example, in all the classes of Eq.~(\ref{list.P_O}), 
possible sets of $(k, k', l)$ are $(0, 0, 0), \, (1, 0, 0), \, (0, 1, 0), \, 
(0, 0, 1), \, (2, 0, 0)$, $(1, 1, 0), \,(1, 0, 1), \,(0, 2, 0), \,(0, 1, 1)$,  
and $(0, 0, 2)$, and the corresponding $P^{L,n_0}_l(m)$ turns out to be simple. 
For $l=0$, $P^{L,0}_0(m)=1$. 
For $l=1$,  $P^{L,0}_1(m)=2m+L$ ($L \geqq 1$) and 
$P^{L,1}_1(m)=2$ ($L=0$). 
For $l=2$,  
$P^{L,0}_2(m)=(2m+L-1)(2m+L)$ ($L \geqq 2$) and $P^{L,1}_2(m)=2(2m+2L+1)$ ($L=0,1$). Once $P^{L,n_0}_l(z\frac{d}{dz})$ is given, its action on 
$\gamma_{K, K', L}(z)$ is performed by using  
\begin{align}
z\frac{d}{dz}\gamma_{K,K',L}(z)
=\big(L+\frac{1}{2}\big)\big[\gamma_{K,K',L-1}(z)-
\gamma_{K,K',L}(z)\big],
\end{align}
which is derived from the well-known formulas involving the GHF. 
Table~\ref{value.F.z=1} tabulates $F^{KKL}_{kk'l}(1)$ for the above
cases. 

With $C$ set to the unit matrix in Eqs.~(\ref{me.xcx}) and (\ref{me.xcx**2}) and using 
Table~\ref{value.F.z=1}, the 
expectation values of $\rho^2$ and $(\rho^2-\langle \rho^2 \rangle)^2$ 
are given by ($\kappa=2K+L$)
\begin{align}
&\langle f^{uA}_{KLM} | \rho^2 | f^{uA}_{KLM} \rangle \equiv \langle \rho^2 \rangle
=\frac{3}{2}{\rm Tr}A^{-1}+
\kappa \frac{\widetilde{u}A^{-2}u}{\widetilde{u}A^{-1}u}, 
\label{xAx}
\\
&\langle f^{uA}_{KLM} | (\rho^2-\langle \rho^2 \rangle)^2 | f^{uA}_{KLM} \rangle 
\notag \\
&\qquad \quad =\frac{3}{2}{\rm Tr}A^{-2}+
2\kappa \frac{\widetilde{u}A^{-3}u}{\widetilde{u}A^{-1}u}-\kappa\Big(\frac{\widetilde{u}A^{-2}u}{\widetilde{u}A^{-1}u}\Big)^2.
\label{xAx**2}
\end{align}
The $\sigma$ value of Eq.~(\ref{stand.dev}) is readily obtained for $f^{uA}_{KLM}$.

\subsection{Hamiltonian matrix element}
\label{H.m.e}

We show how to calculate the matrix element of $H_{\Omega}$. 
First we note that the relative distance vector, ${\bm r}_i-{\bm r}_j$, 
is expressed as a combination of $\bm x_k$,
\begin{align}
&{\bm r}_i-{\bm r}_j=\sum_{k=1}^{N-1}({U^{-1}}_{ik}-{U^{-1}}_{jk}){\bm x}_k \equiv 
\widetilde{\omega^{(ij)}}\bm x,
\label{def.omega}
\end{align}
where $\omega^{(ij)}$ is a column vector of dimension $(N-1)$. Its square is 
$({\bm r}_i-{\bm r}_j)^2=\widetilde{\bm x}T^{(ij)}\bm x$, where 
$T^{(ij)}=\omega^{(ij)}\widetilde{\omega^{(ij)}}$ is a symmetric $(N-1)\times (N-1)$ matrix. A Gaussian potential $e^{-a({\bm r}_i-{\bm r}_j)^2}$ is expressed as 
$e^{-a\widetilde{\bm x}{T^{(ij)}}\bm x}$, and its 
matrix element reduces to the overlap~(\ref{ff.overlap}): 
\begin{align}
\langle f^{u'A'}_{K'LM}|e^{-a\widetilde{\bm x}{T^{(ij)}}\bm x}|f^{uA}_{KLM}\rangle 
=G^{u'A':uA}_{K'L:KL}(aT^{(ij)}),
\label{gauss.pot}
\end{align}
where
\begin{align}
&G^{u'A':uA}_{K'L:KL}(T)\notag \\
&={\cal R}^{u' A':A'+T}_{K'L} {\cal R}^{u A:A+T}_{KL}
\langle f^{u' \, A'+T}_{K'LM}|f^{u \, A+T}_{KLM}\rangle
\label{def.G}
\end{align}
with
\begin{align}
{\cal R}^{u A:A'}_{KL}=\frac{{\cal N}^{uA}_{KL}} {{\cal N}^{u\, A'}_{KL}}
=\Big(\frac{{\rm det}A}{{\rm det}A'}\Big)^{\frac{3}{4}}\Big(\frac{\widetilde{u}{A'}^{-1}u}{\widetilde{u}A^{-1}u}\Big)^{\frac{2K+L}{2}}.
\end{align}
The matrix element of three-body force of Gaussian form factor 
can be obtained in a similar way.

The matrix elements of Coulomb and Yukawa potentials 
are obtained by applying the above result~\cite{suzuki08}. For example, 
by expressing the Yukawa potential as 
\begin{align}
\frac{1}{r}e^{-\mu r} =\frac{2}{\sqrt{\pi}}\int_0^{\infty}dt \exp\big(-t^2r^2-\frac{\mu^2}{4t^2}\big),
\end{align}
its matrix element is obtained by a numerical integration of Eq.~(\ref{gauss.pot}) with an appropriate change of the range parameter $a$. 
Equation~(\ref{gauss.pot}) is valid for not only  
$T^{(ij)}$ but  any positive-definite symmetric matrix.  For example,  using  the unit matrix $I$  
we obtain  
\begin{align}
\langle f^{u'A'}_{K'LM}|\frac{1}{\rho^2}|f^{uA}_{KLM}\rangle 
=\int_0^{\infty} dt \, G^{u'A':uA}_{K'L:KL}(tI),
\label{me.rho-2}
\end{align}
which is computed with e.g. Gauss-Laguerre quadrature.

We turn to the hyperangular kinetic energy $T_{\Omega}$. 
We obtain its 
matrix element without expressing $\Lambda^2$ 
in terms of $\Omega$, but in the indirect 
way~\cite{stecher09} that utilizes the identity, $T_{\Omega}=T-T_{\rho}$. 
The matrix elements of $T$ and $T_{\rho}$ are respectively obtained as 
follows. 
As for $T$, we start from 
\begin{align}
&T g(\bm s, A, \bm x)\notag \\
&=\frac{\hbar^2}{2m}(3{\rm Tr}A-\widetilde{\bm s}\bm s+2\widetilde{\bm s}A\bm x
-\widetilde{\bm x}A^2\bm x)g(\bm s, A, \bm x).
\label{T.on.g}
\end{align}
$T_{kk'l}(u'A',uA)$ contributed by each term of Eq.~(\ref{T.on.g}) 
is read from Appendix~\ref{CG.m.e}.  The use of 
Eq.~(\ref{me.vs.ovlme}) leads to 
\begin{align}
&\langle f^{u'A'}_{K'LM}|T|f^{uA}_{KLM}\rangle \notag \\
&=\frac{\hbar^2}{2m}\langle f^{u'A'}_{K'LM} | f^{uA}_{KLM} \rangle 
\Big[\, \frac{3}{2}{\rm Tr}B-\frac{3}{2}{\rm Tr}C_1C \notag \\ 
&\ \  -(\widetilde{u}u-2\widetilde{u}C_1u+\widetilde{u}C_2u)
\frac{1}{\widetilde{u}B^{-1}u}F^{KK'L}_{100}(z)\notag \\
&\ \ -(\widetilde{u'}u'+2\widetilde{u'}C_1u'+\widetilde{u'}C_2u')\frac{1}{\widetilde{u'}B^{-1}u'}F^{KK'L}_{010}(z)\notag \\
&\ \ +(\widetilde{u}u'+\widetilde{u}C_1u'-\widetilde{u'}C_1u-\widetilde{u}C_2u')\frac{1}{\widetilde{u}B^{-1}u'}F^{KK'L}_{001}(z)\Big],
\end{align}
where $C, C_1$, and $C_2$ are the matrices defined by 
\begin{align}
C=\frac{1}{2}(A-A'),\ \ \  C_1=CB^{-1},\ \ \  C_2=B^{-1}C^2B^{-1}.
\end{align}

As for $T_{\rho}$, we use Eq.~(\ref{cg.rho-dep}) to obtain the relation 
\begin{align}
T_{\rho}f^{uA}_{KLM}&=-\frac{\hbar^2}{2m\rho^2}\Big[\kappa^2+(d-2)\kappa \notag \\
&\ \ \ 
-(2\kappa+d)\widetilde{\bm x}A\bm x+(\widetilde{\bm x}A\bm x)^2\Big]f^{uA}_{KLM}.
\end{align}
As in Eq.~(\ref{me.rho-2}), the matrix element of $T_{\rho}$
is obtained by performing the following integration:
\begin{align}
&\langle f^{u'A'}_{K'LM}|T_{\rho}|f^{uA}_{KLM}\rangle \notag \\
&=-\frac{\hbar^2}{2m}\int_0^{\infty}dt \, 
{\cal R}^{u' A':A'+tI}_{K'L} {\cal R}^{u A:A+tI}_{KL}  \notag \\
&\ \ \times \langle f^{u' \, A'+tI}_{K'LM}|\Big[ 
\kappa^2+(d-2)\kappa
-(2\kappa+d)\widetilde{\bm x}A\bm x \notag \\
&\qquad \qquad \qquad +(\widetilde{\bm x}A\bm x)^2 \Big] |
f^{u \, A+tI}_{KLM}\rangle,
\end{align}
where the matrix elements of $\widetilde{\bm x}A\bm x $ and 
$(\widetilde{\bm x}A\bm x)^2 $ are readily available from Eqs.~(\ref{me.xcx}) 
and (\ref{me.xcx**2}). 
Following Refs.~\cite{stecher09,rittenhouse11,rakshit12}, the matrix element of 
$T_{\Omega}$ is given as 
\begin{align}
\langle f^{u'A'}_{K'LM}|T_{\Omega}|f^{uA}_{KLM}\rangle = &\frac{1}{2}\Big(
\langle f^{u'A'}_{K'LM}|T-T_{\rho}|f^{uA}_{KLM}\rangle\notag \\
&+\langle f^{uA}_{KLM}|T-T_{\rho}|f^{u'A'}_{K'LM}\rangle\Big).
\end{align}

All the matrix elements needed to solve Eq.~(\ref{eq.to.be.solved}) are thus 
available in the CG basis functions. The present approach thus reduces the whole 
task to a standard linear algebra of matrices in place of the coupled 
differential equation commonly used in the hyperspherical approach.

\subsection{Permutation symmetry}

The permutation symmetry for identical particles has to be 
imposed on the wave function. Its incorporation  
in the CG is very easy~\cite{book,suzuki02,suzuki17}. 

The permutation $P$ induces the coordinate transformation: ${\bm x}  \to T_P \bm x$, where 
the $(N-1)\times(N-1)$ matrix $T_P$ is easily determined. Since 
$P$ just rearranges the labels of 
$\bm r_i$, $\rho^2$ remains unchanged (see Eqs.~(\ref{def.rho1}) and (\ref{def.rho})):  $\rho^2=\widetilde{\bm x}\bm x \to \widetilde{T_P \bm x }T_P\bm x
=\widetilde{\bm x}\widetilde{T_P}T_P\bm x=\widetilde{\bm x}\bm x$, concluding  
\begin{align}
\widetilde{T_P}T_P=I.
\end{align} 
The CG acted by $P$ transforms to 
\begin{align}
Pf^{uA}_{KLM}(\bm x)&={\cal N}^{uA}_{KL}|\widetilde{u_P}\bm x|^{2K+L}
Y_{LM}(\widehat{\widetilde{u_P}\bm x})e^{-\frac{1}{2}\widetilde{\bm x}A_P\bm x}\notag \\
&=\frac{{\cal N}^{uA}_{KL}}{{\cal N}^{u_P A_P}_{KL}}f^{u_P A_P}_{KLM}(\bm x),
\end{align}
where 
\begin{align}
u_P=\widetilde{T_P}u,\ \ \ A_P=\widetilde{T_P}AT_P.
\label{uA.vs.P}
\end{align}
Since ${\rm det}A_P={\rm det}A$ and $\widetilde{u_P}A_P^{\,-1}u_P=\widetilde{u}A^{-1}u$, Eq.~(\ref{normalization.const}) confirms 
${\cal N}^{u_PA_P}_{KL}={\cal N}^{uA}_{KL}$, establishing 
\begin{align}
Pf^{uA}_{KLM}(\bm x)=f^{u_P A_P}_{KLM}(\bm x).
\label{perm.on.cg}
\end{align}

The CG keeps its functional form under the permutation, and 
its effect results in simply changing the CG parameters, $u$ and $A$, as in Eq.~(\ref{uA.vs.P}).  
The basis function $\Phi^{uA}_{KLM}$ that is constructed from 
$f^{uA}_{KLM}(\bm x)$ and satisfies the symmetry requirement is given by 
\begin{align}
\Phi^{uA}_{KLM}=\sum_{P}\epsilon_P f^{u_P A_P}_{KLM}(\bm x),
\label{sym.wf}
\end{align}
where $\epsilon_P$ is the phase of $P$.

\section{Test of three-$\alpha$ system}
\label{3alpha.example}

In order to learn similarity to and dissimilarity from the usual adiabatic 
channel energy, we use the same Hamiltonian as that of Ref.~\cite{suno16}. 
The mass of the $\alpha$ particle is  $\hbar^2/m$=10.5254408\,
MeV\,fm$^2$, and the charge constant is $e^2$= 1.4399644\,MeV\,fm. 
The two-body potential $V_{\alpha \alpha}(r)$ consists of a modified 
Ali-Bodmer potential~\cite{ali66} and the Coulomb potential:
\begin{align}
V_{\alpha \alpha}(r)&=125\,e^{-r^2/1.53^2}-30.18\,e^{-r^2/2.85^2}\notag \\
&+\frac{4e^2}{r}{\rm erf}\,(0.60141r),
\end{align}
where the length and energy are given in units of fm and MeV, respectively. 
The three-body potential is chosen to be a hyperscalar potential,  
\begin{align}
V_{\alpha \alpha \alpha}=v_3\,e^{-a_3\widetilde{\bm x}\bm x},
\end{align}
where the range parameter $a_3$ is 
$\sqrt{3}/R_3^2$ with $R_3$=2.58\,fm, and 
the potential strength $v_3$ is $L$-dependent:  It is $-151.737$\,MeV for 
$L=0$ to reproduce the Hoyle resonance energy, and $-179.463$\,MeV for 
$L=2$ to fit the lowest $2^+$ state energy of $^{12}$C.

\subsection{Specification of correlated-Gaussian parameters}

We use $\Phi^{uA}_{KLM}$, Eq.~(\ref{sym.wf}), as the basis functions 
$\Phi_{R_i,l}$. The label $l$ stands for $K, u$, and $A$.  
$u$ contains just 1 parameter, assuming that $u$ is normalized:  
$u_1=\sin \zeta,\,  u_2=\cos \zeta\, (0 \leqq  \zeta < \pi)$. 
$\zeta$ is discretized by $M_{\zeta}$ meshes. 
The matrix $A$ for three-body system contains 3 parameters, $A_{11}, A_{12}(=A_{21}), A_{22}$. 
It may be prescribed with three parameters 
$(d_{12},  d_{23}, d_{13})$ as 
\begin{align}
\widetilde{\bm x}A\bm x=\sum_{j>i=1}^3\frac{1}{{d_{ij}}^{\!2}}(\bm r_i-\bm r_j)^2.
\end{align}
Roughly speaking, $d_{ij}$ controls the distance between particles $i$ and $j$. 
In analogy to the prescription used in Refs.~\cite{suno15,suno16}, we specify 
$d_{ij}$  by two angles $\theta\ (0\leqq \!\theta<\! \pi/2)$ and 
$\phi\ (0 \leqq \phi \leqq \pi)$ that define the `shape' of three particles:  
\begin{align}
&{d_{12}}^{\!2}={\bar d}^{\, 2} \Big[1+\sin \theta \cos\big(\phi+\frac{2}{3}\pi \big)\Big]
\equiv \frac{{\bar d}^{\,2}} {\lambda_+},\notag \\
&{d_{23}}^{\!2}={\bar d}^{\,2}[1+\sin \theta \cos \phi ]\equiv \frac{{\bar d}^{\,2}}{\lambda_0},\notag \\
&{d_{13}}^{\!2}={\bar d}^{\,2}\Big[1+\sin \theta \cos\big(\phi-\frac{2}{3}\pi \big)\Big]
\equiv \frac{{\bar d}^{\,2}}{ \lambda_-}.
\end{align}
$\theta\!=\!0$ and $\theta\!=\!\pi/2$ correspond to equilateral triangle and collinear configurations, respectively. Since we have three identical particles, 
the range of $\phi$ can be restricted to $[0, \pi/3]$. We discretize $\theta$ and $\phi$ by $M_{\theta}$ and $M_{\phi}$ meshes. 
The matrix $A$ reads as $A=A_0/{\bar d}^{\,2}$, where  
\begin{align}
A_0=\Big(
\begin{array}{ccc}
2\lambda_++\frac{1}{2}(\lambda_0+\lambda_-) && -\frac{\sqrt{3}}{2}(\lambda_0-\lambda_-) \\
-\frac{\sqrt{3}}{2}(\lambda_0-\lambda_-)  && \frac{3}{2}(\lambda_0+\lambda_-) \\
\end{array}
\Big),
\end{align}
and ${\bar d}^{\,2}$ is determined from the 
constraint~(\ref{rho2.constraint}).

\subsection{Results}
  
As defined in Eq.~(\ref{rho2.constraint}), $R$ is a $c$-number representing 
$\sqrt{\langle \rho^2 \rangle}$. In Refs.~\cite{suno15,suno16}, 
$R$ stands for both the hyperradius operator and its value, although 
the hyperradius there corresponds to $3^{1/4}\rho$ 
of the present paper. To avoid confusion,  
we employ a point-$\alpha$ root-mean-square (rms) radius $R_{\rm rms}$ as a length scale, 
\begin{align}
R_{\rm rms}=\sqrt{\frac{\langle \rho^2 \rangle}{3}}, 
\end{align}
which is computed as $3^{-\frac{3}{4}}R$ from $R$ in~\cite{suno15,suno16}.

\begin{figure}
\begin{center}
\includegraphics[scale=0.3]{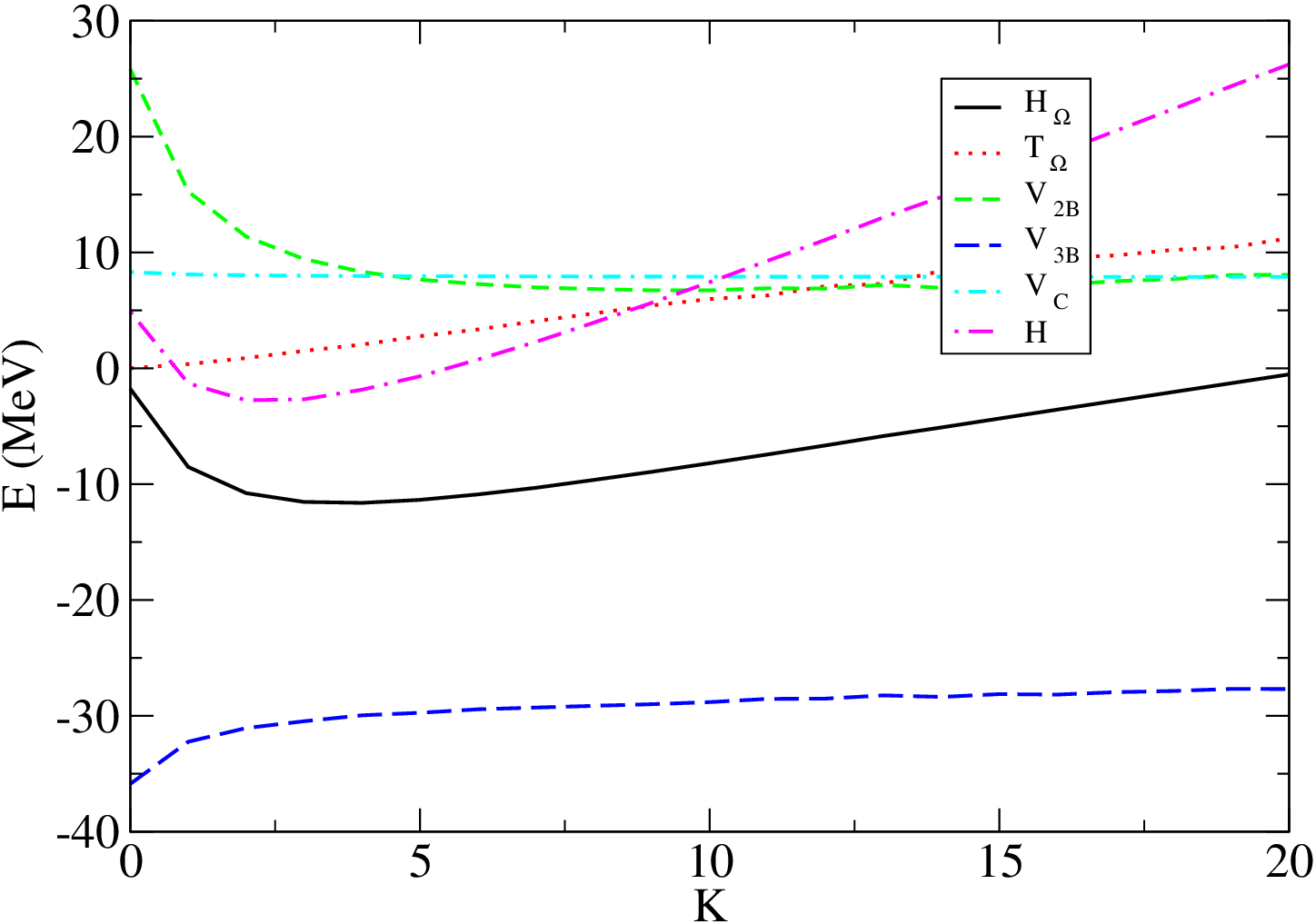}
\caption{(Color online) Minimum  expectation value of $H_{\Omega}$ calculated by 
a single configuration, $\Phi^{uA}_{KL=0\,M=0}$, Eq.~(\ref{sym.wf}), as a function of $K$. The minimum is searched for by varying $\theta, \phi, \zeta$ on the meshes discretized with  $M_{\theta}=30$, $M_{\phi}=20$,  
and $M_{\zeta}=30$ under the constraint that $R_{\rm rms}$ is kept to 1.54\,fm. 
The contributions of $T_{\Omega}$, the nuclear potentials (two-body $V_{\rm 2B}$, 
and three-body $V_{\rm 3B}$) as well as the Coulomb potential ($V_{\rm C}$) 
to the minimum energy are also drawn. The curve denoted $H$ is the variation 
of the total energy. }
\label{fig1}
\end{center}
\end{figure}

\begin{figure}
\begin{center}
\includegraphics[scale=0.3]{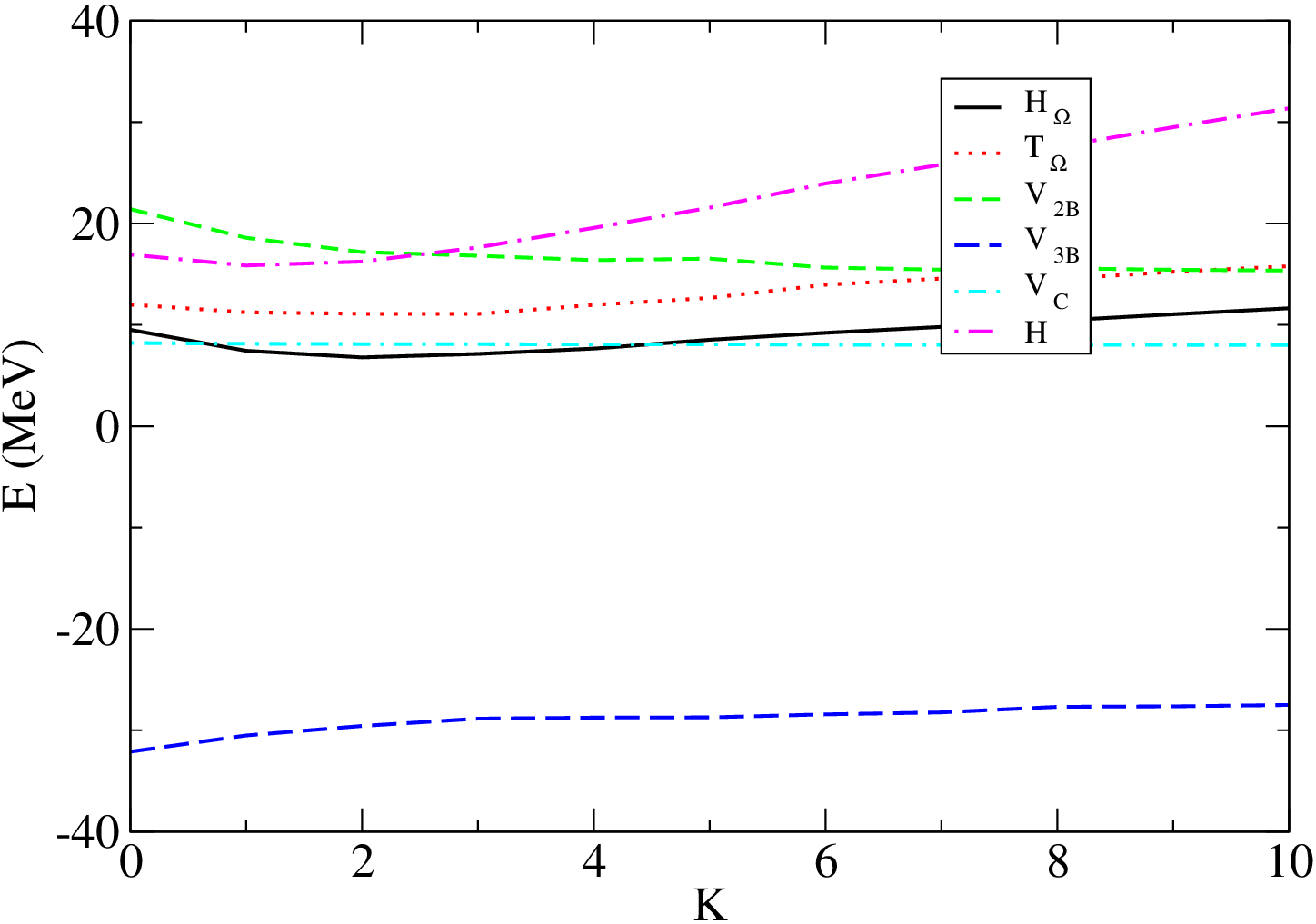}
\caption{(Color online) The same as Fig. 1 but for $L^{\pi}=2^+$. }
\label{fig2}
\end{center}
\end{figure}

Figure~\ref{fig1} displays $K$-dependence of the  
minimum  expectation value of $H_{\Omega}$ 
calculated by a single configuration with $L^{\pi}=0^+$. 
For each $K$, $u$ and $A$ are varied 
on the meshes discretized with $M_{\zeta}, M_{\theta}$, and $M_{\phi}$, 
subject to $R_{\rm rms}=1.54$\,fm. 
The minimum of the adiabatic channel energy occurs around 
that rms value~\cite{suno16}. 
The minimum of the curve is $-11.62$\,MeV at $K=4$ and gradually increases with 
$K$. The contribution of $T_{\Omega}$ to the minimum expectation value 
increases as $K$ increases, 
while the sum of the potentials,  
$V_{\rm 2B}+V_{\rm 3B}+V_{\rm C}$, shows a moderate change 
for $K \geqq 2$, probably because it is determined mainly by 
the global size of the system. The curve labeled $H$ is the sum of the 
minimum energy of $H_{\Omega}$ and the expectation value of $T_{\rho}$, that is, 
the total energy, calculated by the optimal configuration.    
The expectation value of $T_{\rho}$ 
increases from 7 to 27\,MeV as $K$ increases from 0 to 20. 
Figure~\ref{fig2} is the same as Fig.~\ref{fig1} but for $L^{\pi}=2^+$. 
The configuration is again constrained to satisfy $R_{\rm rms}=1.54$\,fm.  
The minimum energy of $H_{\Omega}$ is 6.79 MeV at $K=2$.

\begin{table}
\caption{The properties of the single configuration used 
in Figs.~\ref{fig1} and \ref{fig2}.  $\sigma$ is 
the standard deviation and Overlap is the overlap integral 
with $K=4$ ($L=0$) or $K=2$ ($L=2$) basis function. 
 } 
\label{Fig.1data}
\begin{tabular}{ccccccccc}
\hline\hline
   &&&  \quad $L=0$ &    &&& \quad $L=2$ &  \\
\hline
$K$ &&& $\sigma$ & Overlap &&& $\sigma$ & Overlap \\
\hline 
0 &&& 0.577 & 0.923 &&& 0.477 & 0.977 \\
1 &&& 0.480 & 0.973 &&& 0.431 & 0.995 \\
2 &&& 0.450 & 0.991 &&& 0.400 & 1.000 \\
3 &&& 0.435 & 0.998 &&& 0.374 & 0.996 \\
4 &&& 0.419 & 1.000 &&& 0.378 & 0.985 \\
5 &&& 0.413 & 0.999 &&& 0.381 & 0.978 \\
6 &&& 0.403 & 0.996 &&& 0.373 & 0.952 \\
7 &&& 0.399 & 0.992 &&& 0.365 & 0.941 \\
8 &&& 0.394 & 0.987 &&& 0.338 & 0.929 \\
9 &&& 0.389 & 0.981 &&& 0.338 & 0.919 \\
10&&& 0.382 & 0.975 &&& 0.332 & 0.909 \\
11&&& 0.370 & 0.967 &&& & \\
12&&& 0.370 & 0.962 &&& & \\
13&&& 0.358 & 0.952 &&& & \\
14&&& 0.364 & 0.950 &&& & \\
15&&& 0.354 & 0.940 &&& & \\
16&&& 0.356 & 0.936 &&& & \\
17&&& 0.346 & 0.927 &&& & \\
18&&& 0.342 & 0.919 &&& & \\
19&&& 0.334 & 0.909 &&& & \\
20&&& 0.335 & 0.905 &&& & \\
\hline\hline
\end{tabular}
\end{table}

Table~\ref{Fig.1data} lists some properties of the single configuration used 
in Figs.~\ref{fig1} and \ref{fig2}. It is noted that the  
standard deviation $\sigma$ decreases as $K$ increases. The configuration with $K=4$ 
giving the energy minimum for $L=0$ has $\sigma=0.419$. Roughly speaking, 
this $\sigma$ value corresponds to the degree of localization,  
$(\sqrt[4]{1+\sigma^2}-1)R_{\rm rms} \approx 0.064$\,fm around $R_{\rm rms}$. 
If we want to use more localized configurations, we have to increase $K$. 
Since the overlap with the $K=4$ configuration decreases very slowly as listed in 
Overlap column, the 
energy loss may not be very large. In $L=2$ case, the $K$ dependence of 
$\sigma$ and the overlap integral appears to decrease faster than the $L=0$ case.

\begin{figure}
\begin{center}
\includegraphics[scale=0.3]{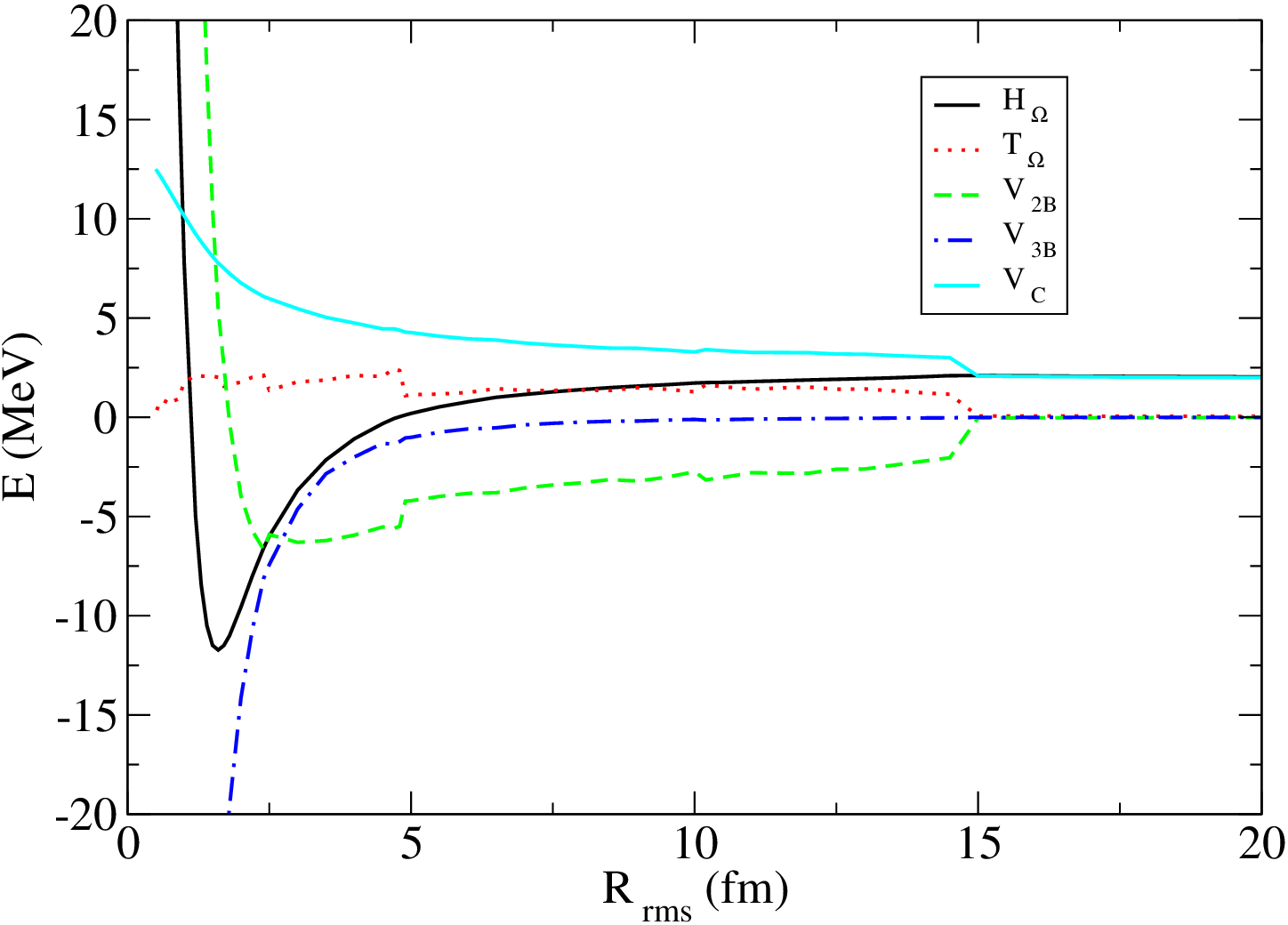}
\vskip 0.4cm
\includegraphics[scale=0.3]{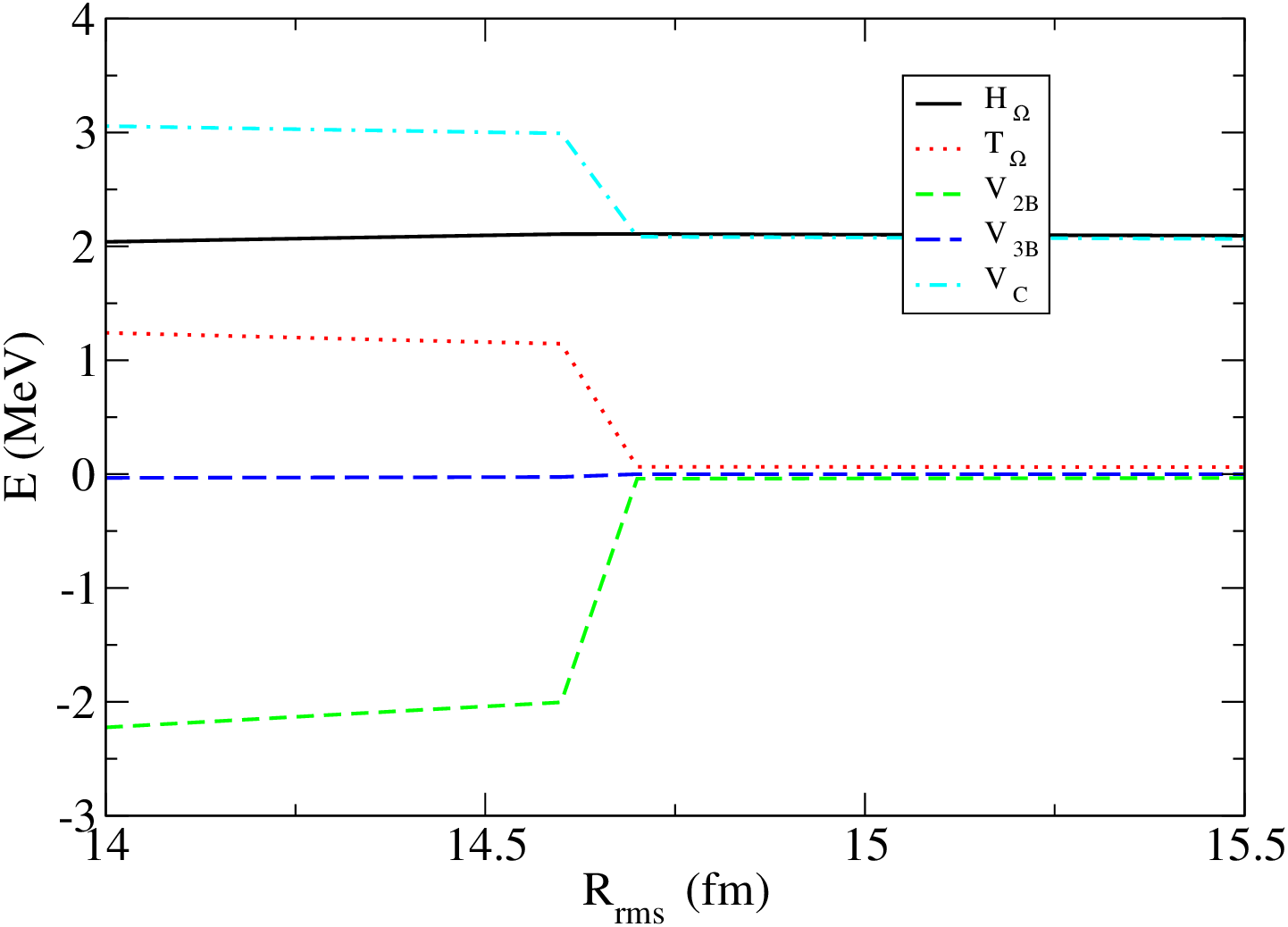}
\caption{(Color online) (a) Minimum expectation value of $H_{\Omega}$ 
calculated by a single configuration, $\Phi^{uA}_{K\, L=0\,M=0}$, as a 
function of $R_{\rm rms}$. 
The minimum is searched for by varying $K$ as well as $u$ and 
$A$ that are discretized on the meshes with $M_{\theta}=30$, $M_{\phi}=21$, and 
$M_{\zeta}=45$. The contributions of $T_{\Omega}$, $V_{\rm 2B}$, 
$V_{\rm 3B}$, and $V_{\rm C}$ to the minimum energy are also drawn. 
(b) An enlarged figure of (a) in $R_{\rm rms}=14.0-15.5$\,fm. }
\label{fig3}
\end{center}
\end{figure}

Figure~\ref{fig3}(a) plots the 
minimum expectation value of $H_{\Omega}$ for $L^{\pi}=0^+$
as a function of $R_{\rm rms}$. 
The minimum energy is obtained by a single configuration determined 
similarly to the case of 
Fig.~\ref{fig1}, but with slightly finer meshes.  We learn 
how each term of $H_{\Omega}$ responds to 
the expansion of the system as  $R_{\rm rms}$ increases. 
The $H_{\Omega}$ curve shows a minimum 
around $R_{\rm rms}$=1.6\,fm, and reaches a broad tiny peak 
at 14.6-14.7\,fm, where  
the contribution of each piece of $H_{\Omega}$ displays a sudden change as 
magnified in Fig.~\ref{fig3}(b).  
Before the peak, $V_{\rm C}$, $T_{\Omega}$, and $V_{\rm 2B}$ are main 
contributors to the $H_{\Omega}$ curve, whereas after the peak both 
contributions 
of $T_{\Omega}$ and $V_{\rm 2B}$ get small and $V_{\rm C}$ plays a dominant role. 
Note, however, that the contribution of $V_{\rm 2B}$ persists up 
to large distances beyond 14\,fm, despite the fact that the range of 
$V_{\alpha \alpha}$ is much shorter than that value. This long-range effect 
is due to the $\alpha \alpha$ resonance. 
  
Although the minimum expectation value of $H_{\Omega}$ changes smoothly with 
$R_{\rm rms}$ as seen in Fig.~\ref{fig3}(a), the contributions of $T_{\Omega}$ 
and $V_{\rm 2B}$ show some kinks, especially when $R_{\rm rms}$ changes 
from 1.6 to 1.7, 2.4 to 2.5, and 4.8 to 4.9 fm. At these points 
the optimal $K$ value also changes as follows: 
4$\to$3, 3$\to$ 2, and 2$\to$1, respectively. However, 
the minimum expectation value of $H_{\Omega}$ is often 
not very sensitive to the change of $K$ but several $K$ configurations 
give almost equal results, whereas the contribution of $T_{\Omega}$ seems to 
be more sensitive to $K$. This is understood from the degree of 
localization of the CG. In fact, the   
$\sigma$ value of $\Phi^{uA}_{K L M}$ decreases with increasing $K$, 
and hence the contribution of $T_{\Omega}$ tends to increase.

Now we mix various configurations to solve Eq.~(\ref{eig.channel}).  
The constrained equation is solved at 
the following four points: $R_{\rm rms}$=1.6, 2.5, 5.0, 14.5 fm. 
The lowest adiabatic channel energy of Ref.~\cite{suno16} exhibits 
different character at these points, a steep slope close to the
minimum, and a broad plateau close to the three-$\alpha$ threshold. 
The CG basis functions are generated by including different 
$K, \theta, \phi$, and $\zeta$ parameters. $K$ is 
tested up to 20. The mesh points are 
discretized with $M_{\theta}=30$, $M_{\phi}=21$, and $M_{\zeta}=45$. 
To avoid possible linear-dependence of the generated basis 
functions, we exclude any basis function that has overlap of 
more than 0.95 with other basis functions. 
We also exclude any configuration whose 
expectation value of $H_{\Omega}$ is larger than a cut-off energy, $E_c$. 
The value of $E_c$ is a bit arbitrary, and it is taken fairly large compared to 
the expected lowest adiabatic channel energy. 
The actual basis size is around 250. 
Note that the basis functions all have $\langle \rho^2 \rangle=R^2$ but they 
have different $\sigma$ values within $\sigma \leqq 1$.

\begin{figure*}
\begin{subfigure}[b]{0.45\textwidth} 
{\centering \includegraphics[width=0.9\textwidth]{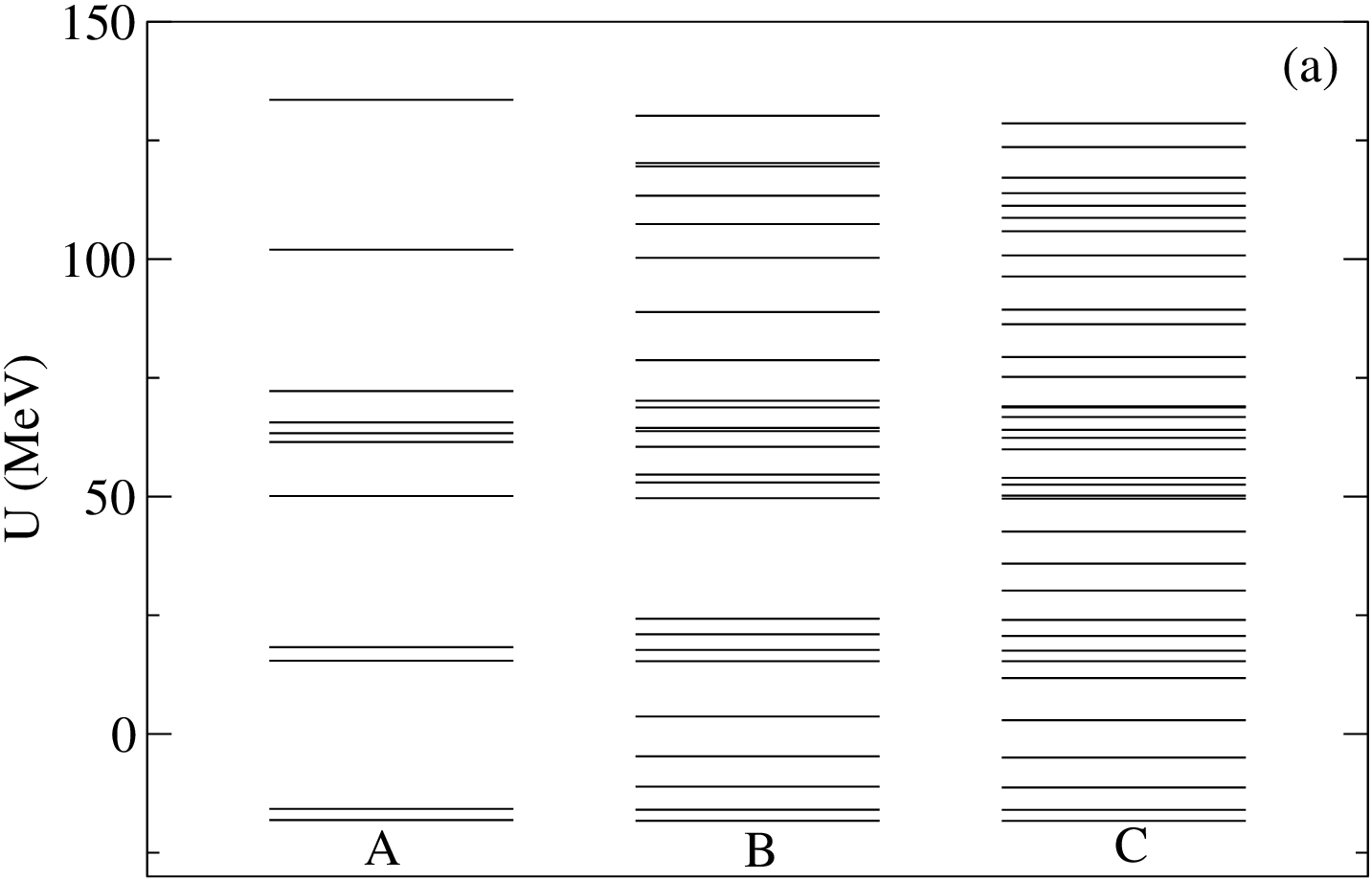}}
\end{subfigure}
\hfill
\begin{subfigure}[b]{0.45\textwidth}  
{\centering \includegraphics[width=0.9\textwidth]{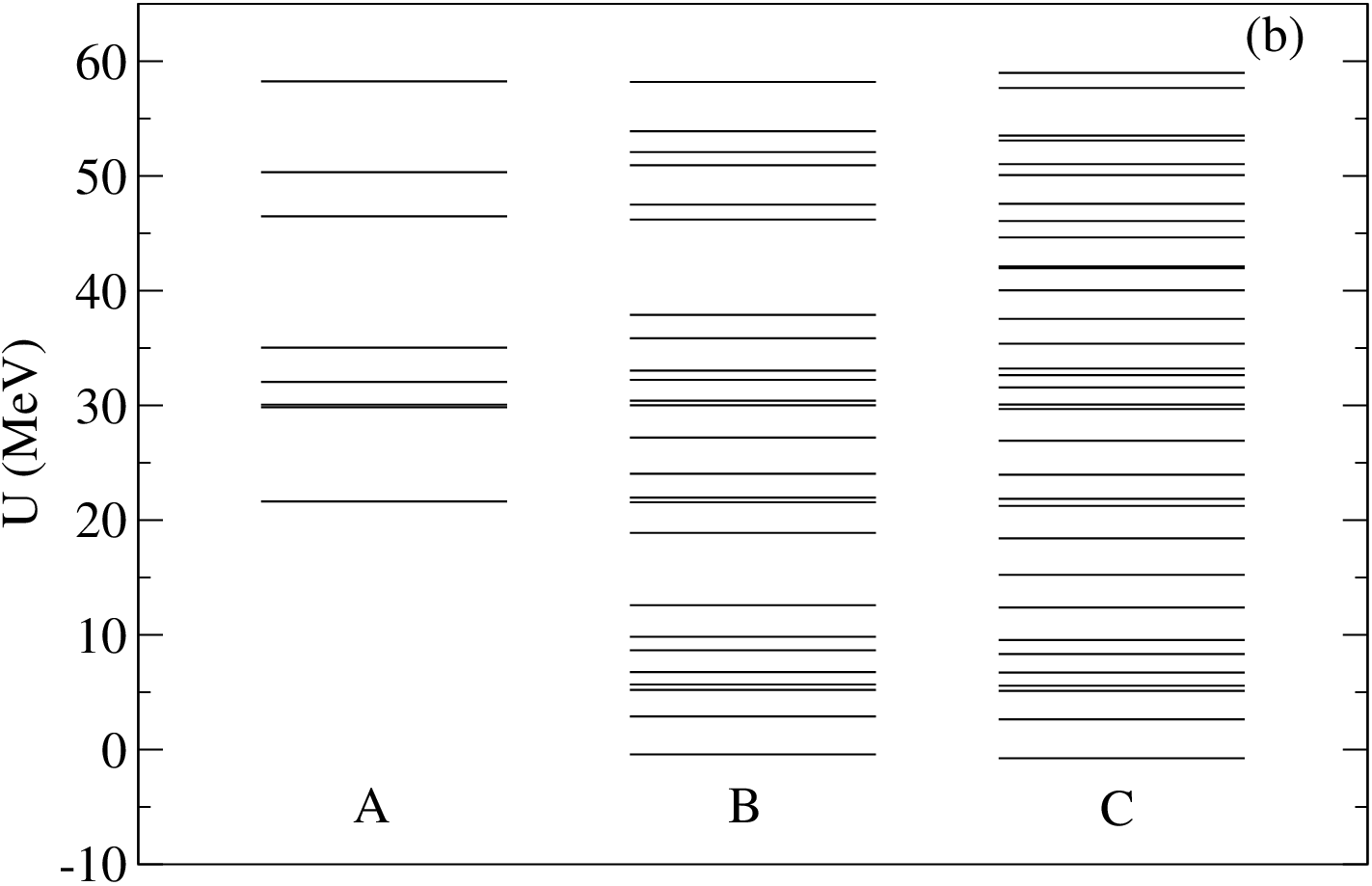}}
\end{subfigure}
\vskip 0.4cm
\vskip\baselineskip
\begin{subfigure}[b]{0.45\textwidth}  
{\centering \includegraphics[width=0.9\textwidth]{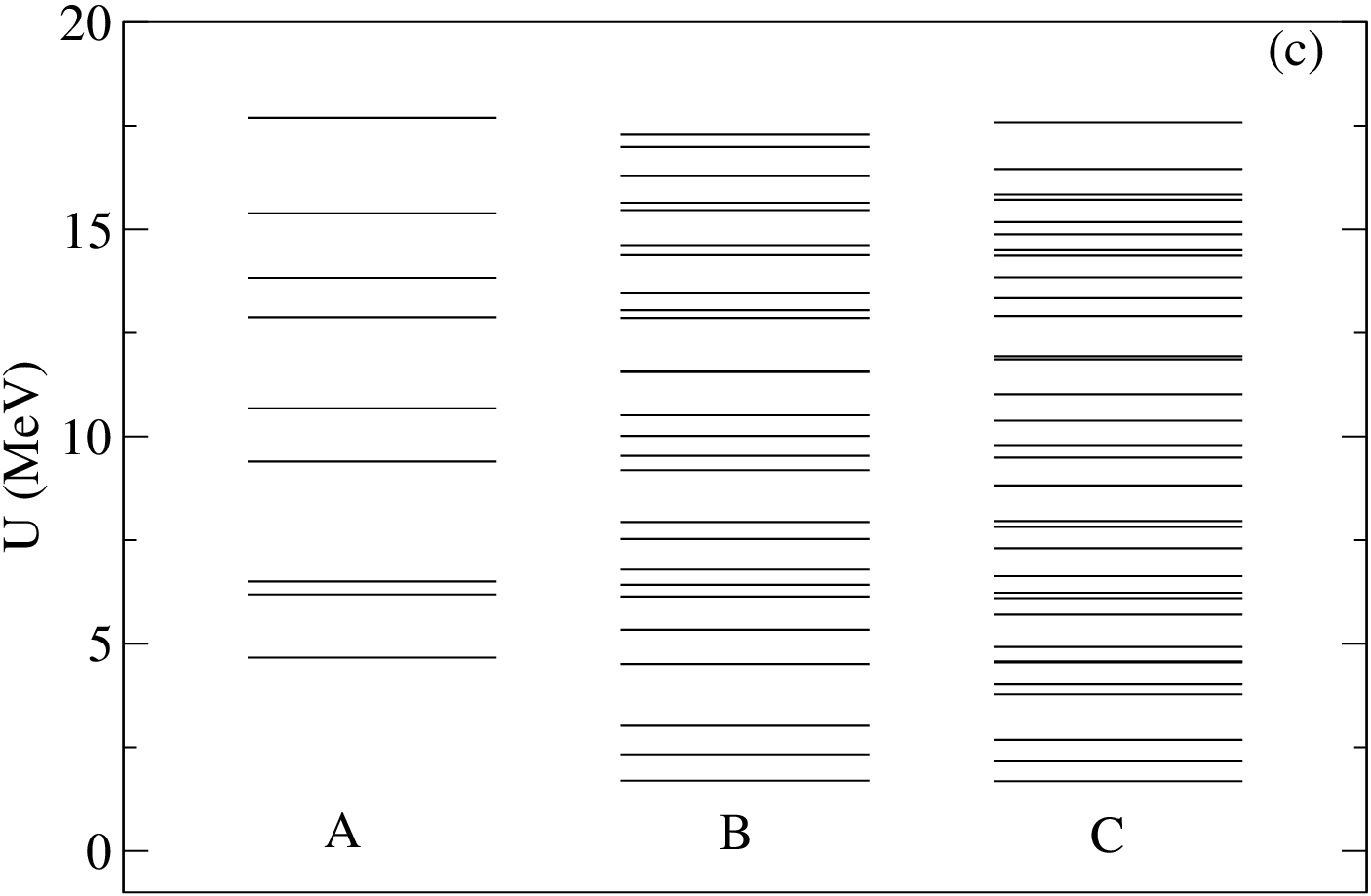}}
\end{subfigure}
\hfill
\begin{subfigure}[b]{0.45\textwidth}  
{\centering \includegraphics[width=0.9\textwidth]{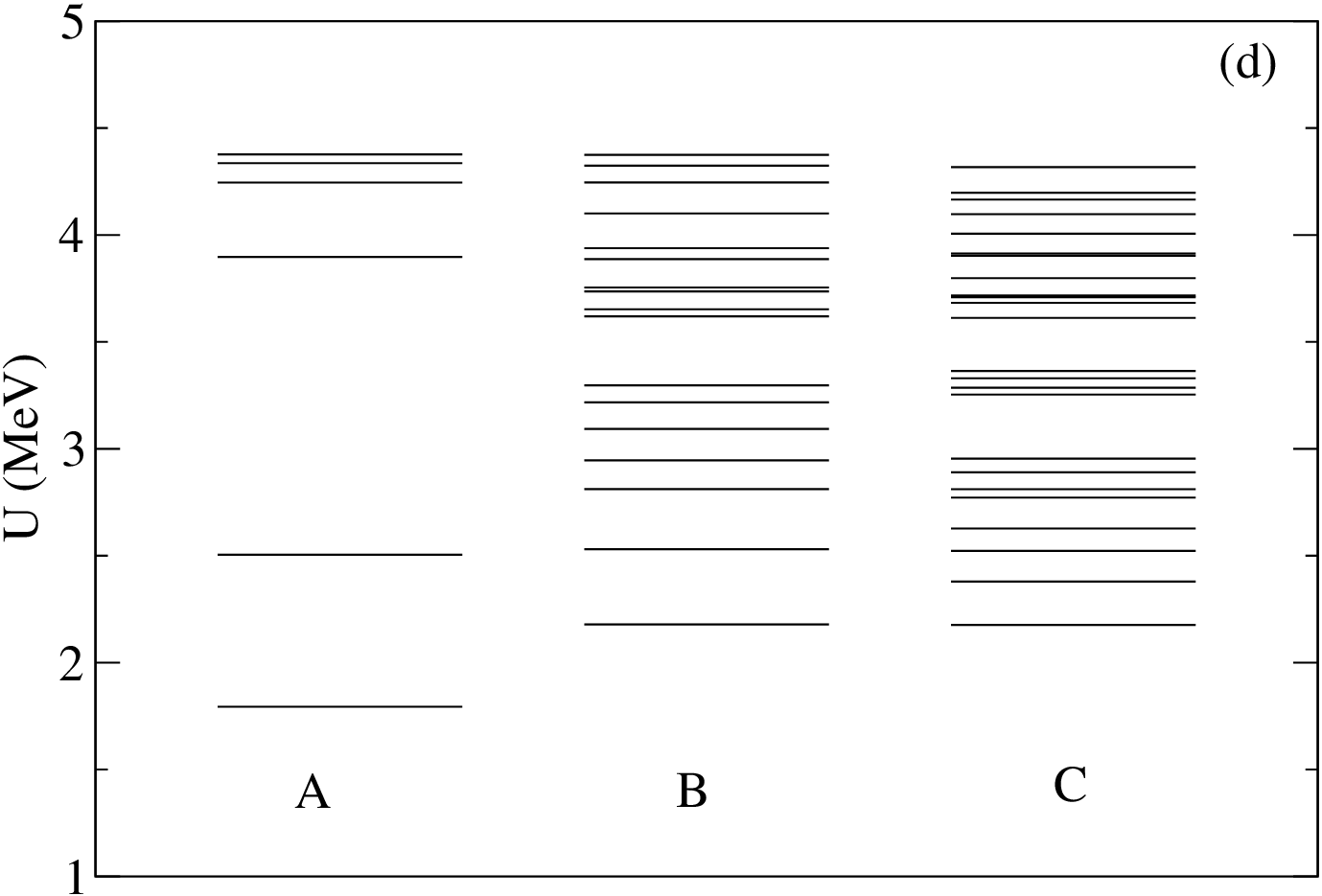}}
\end{subfigure}
\caption{(a) Adiabatic channel energies $U$ at $R_{\rm rms}=1.6$ fm. The basis 
functions are restricted to have the standard deviation $\sigma \leqq 0.5$ 
in case A, 
and $\sigma \leqq 0.75$ in case B, while all the basis functions selected 
are allowed in case C. See the text for detail. Figures (b), (c), and (d) 
are the same as (a) but for $R_{\rm rms}=2.5,\, 5.0$, and $14.5$ fm,
respectively.}   
\label{fig4}
\end{figure*}
			       
In order to see how a spectrum of the adiabatic channel 
energies changes as the basis size increases, we have tested three 
calculations: Case A adopts only those basis functions with 
$\sigma \leqq 0.5$, case B those with $\sigma \leqq 0.75$, 
and case C is a full basis calculation. In each case we calculate the 
$\sigma$ value of $\Phi^{JM\pi}_{R\nu}$ and if it is not larger than $\sigma$ 
that characterizes each case, we accept that $\Phi^{JM\pi}_{R\nu}$ 
as a solution, otherwise it is discarded. Figure~\ref{fig4} plots 
the adiabatic channel energies in each case 
at four $R_{\rm rms}$ radii. The solution of the constrained 
equations,~(\ref{eig.channel}) and (\ref{rho2.constr}), appears to be obtained 
stably. With the increase of the basis size from case A to case C, the density 
of the adiabatic channel energies considerably increases. Note the different energy 
scale in Fig.~\ref{fig4}(a) to~\ref{fig4}(d). 

It is interesting to compare 
the present adiabatic channel energies with those of Ref.~\cite{suno16}. 
The latter uses basis functions quite different from ours: 
At each $\rho$, the channel wave function is expanded in terms of a combination of 
the product of the Wigner $D$ function and fifth-order basis splines 
for the two hyperangles. It includes no $\rho$-dependence. In contrast to this, 
our channel wave function has finite $\rho$-dependence, and therefore receives influence 
from the adiabatic Hamiltonian at nearby $\rho$ values. Thus both energies 
need not be necessarily the same but a comparison may indicate characteristics 
of different types of calculations. 
Three lowest adiabatic channel energies in MeV obtained 
in~\cite{suno16} are $-17.5,\, 15.9,\, 49.7$ at $R_{\rm rms}=1.6$ fm, 
$-4.38,\, 5.55,\, 21.0$ at 2.5 fm, $0.86,\, 3.81,\, 5.38$ at 
5.0 fm, and $0.46,\, 0.83,\, 1.16$ at 14.5 fm, respectively. 
Our corresponding energies are $-18.3,\, -16.0\, -11.3$ at 1.6 fm, 
$-0.75,\, 2.63,\, 5.12$ at 2.5 fm, $1.68,\, 2.16,\, 2.68$ at 5.0 fm, and 
$2.18,\, 2.38,\, 2.52$ at 14.5 fm. The energy spacing of our calculation 
is much narrower than that of Ref.~\cite{suno16}. Which of the two calculations 
gives lower value for the lowest channel energy seems to depend on $R_{\rm rms}$. 
At $R_{\rm rms}=2.5$ fm, the calculation of case C actually gives four energies 
that are lower than $-0.75$ MeV. Since their $\sigma$ values are larger than 1, 
they are not drawn in Fig.~\ref{fig4}(b). Note, however, that the 
highest one among the 
four is predicted to be $-4.09$ MeV with $\sigma=1.008$. A calculation with a 
slightly larger basis set or an optimized basis set would easily predict the 
lowest adiabatic channel 
energy around $-4.4$ MeV. The same thing applies to Fig.~\ref{fig4}(d). The lowest 
energy of case C is higher than that of case A. A few solutions of 
case C are, however, lower than the lowest adiabatic channel energy of case A, but 
they are not shown because their $\sigma$ values are larger than 1. One of 
them is located at 1.79 MeV and has $\sigma=1.08$. It is 
still considerably higher than the lowest energy, $0.46$ MeV, of Ref.~\cite{suno16}.

\section{Summary}
\label{summary}

We have formulated hyperspherical calculations using the flexibility of 
the correlated Gaussians. Differently from conventional hyperspherical methods, 
the channel wave function and the adiabatic channel energy are defined 
by solving the hyperradius-constrained eigenvalue equation of the 
adiabatic Hamiltonian. This approach enables us to perform standard
configuration interaction calculations. 

This work takes a non-conventional venue by allowing the 
spread of the value of the hyperradius for a given basis
function. While in previous hyperspherical calculations, e.g. 
in~\cite{suno16}, the basis functions belong to a given
hyperradius, in the present work ``the basis functions are localized'', 
that is, the hyperradii of the basis functions reside in a narrow
region around a predefined hyperradius. This approach has its advantages 
and disadvantages. A slight disadvantage is that
one can not define a sharp hyperradius, so that the direct comparison to 
conventional calculations is not simple, whereas an advantage is
that the basis functions directly couple the neighboring regions which 
may help to resolve complicated dynamical processes. 
A further advantage is the easier access to larger systems. 

The present formulation is expected to have many applications. 
As an example, we just mention one problem, the fragmentation or decay 
of a nucleus into several $\alpha$ particles at large distances, as 
discussed in~\cite{girod13,royer15,royer14}. 
The approaches employed there have limitations in taking into 
account important effects such as couplings with other 
configurations, the angular momentum dependence of the adiabatic potential, 
and the removal of spurious center-of-mass excitations. Since the issue 
is exactly concerned with how the system evolves as it expands, 
it is worthwhile attempting at resolving those problems in the 
hyperspherical approach.

\acknowledgements

We are grateful to H. Suno for his interest in this work and for sending us his 
calculated results. We thank 
D. Blume, Q. Guan, T. Morishita, and I. Shimamura for useful discussions 
at the early stage of the work. 

\appendix

\section{Solving a constrained eigenvalue problem}
\label{eigenvalue.constraint}

The aim of this appendix is to 
solve a problem of obtaining the 
eigenvalues and corresponding eigenfunctions of a Hermitian 
operator $H$ with a constraint. It is formulated as follows: 
Let $Q$ be a positive-definite Hermitian operator, and  
let ($\phi_1, \phi_2, \ldots, \phi_n$) be a given set of 
normalized, independent basis functions.  
Obtain, in the space spanned by the set, as many $\Phi$'s possible 
that make the expectation value of $H$ 
\begin{align}
\frac{\langle \Phi| H | \Phi \rangle}{\langle \Phi|\Phi \rangle}
\label{eq.1}
\end{align} 
stationary under the constraint 
\begin{align}
\frac{\langle \Phi| Q | \Phi \rangle}{\langle \Phi|\Phi \rangle}=q,
\label{eq.2}
\end{align}
where $q$ is a positive constant. 
The condition $\langle \phi_i|Q | \phi_i \rangle=q$ 
is assumed for each $\phi_i$ in the text. It may 
not be absolutely necessary, however, although the number of 
solutions may depend on how many basis functions satisfy the condition. 

This type of problem appears in several cases. See, for example, 
Ref.~\cite{flocard73} for the optimization of $\Phi$ and 
Ref.~\cite{kukulin78} for the 
determination of $\Phi$ free from some configurations. 
The present problem has distinct differences from those cases in that 
the available configuration space is preset and several solutions 
are requested if possible.

We construct an orthonormal set, 
$(\psi_1, \psi_2, \ldots, \psi_n)$, that makes $Q$ diagonal.
To do this, we first diagonalize the overlap matrix $(\langle \phi_i|\phi_j \rangle)$: 
\begin{align}
\sum_{j=1}^n \langle \phi_i|\phi_j \rangle u^{(k)}_j=b_ku^{(k)}_i,
\label{eig.overlap}
\end{align}
where $\sum_{i=1}^n u^{(k)}_i u^{(l)}_i=\delta_{kl}$. The basis set $u_k$ defined by 
\begin{align}
u_k=\frac{1}{\sqrt{b_k}}\sum_{i=1}^n u^{(k)}_i\phi_i\ \ \ (k=1,\ldots, n)
\label{func.u}
\end{align} 
is orthonormal, $\langle u_k|u_l\rangle=\delta_{kl}$. 
Next, diagonalizing $Q$ in the set $u_k$, 
\begin{align}
\sum_{j=1}^n \langle u_i |Q| u_j \rangle \psi^{(k)}_j=q_k \psi^{(k)}_i,  
\end{align}
with $\sum_{i=1}^n \psi^{(k)}_i\psi^{(l)}_i=\delta_{kl}$, we construct the  
set $\psi_k$ as 
\begin{align}
\psi_k=\sum_{i=1}^n \psi^{(k)}_i u_i,
\label{func.psi}
\end{align}
which has the desired property, $\langle \psi_k|\psi_l\rangle=\delta_{kl}$,   
and $\langle \psi_k|Q|\psi_l\rangle=q_k\delta_{kl}$. 
It is easy to express $\psi_i$ in terms of the original set $\phi_i$'s. 

We attempt to obtain $\Phi$'s step by step.  Defining 
a Hermitian operator $H'$ with a Lagrange multiplier $\lambda$, 
\begin{align}
H'(\lambda)=H-\lambda (Q-q),
\label{aux.H}
\end{align}
we solve the eigenvalue problem,
\begin{align}
H'(\lambda)\Phi(\lambda)=E'(\lambda)\Phi(\lambda),
\label{eq.3}
\end{align}
using the set $\psi_i$, and calculate the expectation value, 
\begin{align}
F(\lambda)&=\langle \Phi(\lambda)|Q-q|\Phi(\lambda) \rangle \notag \\
&=\langle \Phi(\lambda)|Q|\Phi(\lambda) \rangle -q, 
\label{eq.4}
\end{align}
where $\Phi(\lambda)$ is normalized. Focusing always on 
the lowest-energy solution for any $\lambda$, we vary 
$\lambda$ to find a zero of $F(\lambda)$: $F(\lambda_1)=0$. Then  
$\Phi(\lambda_1)$ satisfies the constraint~(\ref{eq.2}) 
and that is the solution to be found: $\Phi_1=\Phi(\lambda_1)$ with 
the energy $E_1=E'(\lambda_1)$. 

To determine the next solution, we define a 
configuration space of dimension $(n-1)$ by removing $\Phi_1$ from the 
set $(\psi_1, \psi_2, \ldots, \psi_n )$, and follow the above procedure 
to find a successful solution $\Phi_2$. This process continues until 
no new solution is found. 
Clearly $\Phi_i$'s determined in this way are orthogonal to each other. 

We can show that   
$\Phi_i$ and $\Phi_j$ for $i \neq j$ 
have no coupling matrix element 
of $H$ if $\lambda_i \neq \lambda_j$. 
Since both functions are orthogonal, the matrix element of $H$ 
reduces to that of $Q$ as follows:
\begin{align}
\langle \Phi_j |H| \Phi_i\rangle
&=\langle \Phi_j |H'(\lambda_i)+\lambda_i(Q-q)| \Phi_i\rangle \notag \\
&=E'(\lambda_i)\langle \Phi_j |\Phi_i\rangle
+\lambda_i\langle \Phi_j |Q-q| \Phi_i\rangle\notag \\
&=\lambda_i\langle \Phi_j |Q| \Phi_i\rangle.
\end{align}
Because of $\langle \Phi_j |H| \Phi_i\rangle=\langle \Phi_i |H| \Phi_j\rangle^*
=\lambda_j \langle \Phi_i |Q| \Phi_j\rangle^*=\lambda_j 
\langle \Phi_j |Q| \Phi_i\rangle$, it follows that 
\begin{align}
(\lambda_i-\lambda_j)\langle \Phi_j |Q| \Phi_i\rangle=0. 
\end{align}
If $\lambda_i \neq \lambda_j$, $\langle \Phi_j |Q| \Phi_i\rangle$ vanishes 
and consequently $\langle \Phi_j |H| \Phi_i\rangle$ must vanish. 

If $\lambda_i$ and $\lambda_j$ are accidentally equal, the above argument 
does not apply and it is not clear whether or not $H$ has the coupling matrix element.

\section{Examples of correlated-Gaussian matrix element}
\label{CG.m.e}

We show the examples of Eq.~(\ref{basic.integral}) that appear frequently. 
For ${\cal O}(\bm x)$, let us consider the following terms:
\begin{align}
&{\rm (i)}\ \ \ \ {\widetilde{\bm w}}\bm x,\notag \\
&{\rm (ii)}\ \ \  ({\widetilde{\bm w}}\bm x) ({\widetilde{{\bm w}'}}\bm x),\notag \\
&{\rm (iii)}\ \ \  {\widetilde{\bm x}}C\bm x,\notag \\
&{\rm (iv)}\ \ \   ({\widetilde{\bm w}}\bm x)({\widetilde{\bm x}}C\bm x),\notag \\
&{\rm (v)}\ \ \ \  ({\widetilde{\bm x}}C\bm x)({\widetilde{\bm x}}C'\bm x).
\label{type.O}
\end{align} 
Here $\bm w$ and ${\bm w}'$ are  
column vectors of dimension $(N-1)$ whose elements are 3-dimensional vectors, 
and 
$C$ and $C'$ are $(N-1)\times (N-1)$ symmetric matrices, and they are all 
independent of $\bm x$. The integral of Eq.~(\ref{basic.integral}) 
is easily obtained. Corresponding to (i)-(v) classes, ${\cal P_O}$ reads  
\begin{align}
&{\rm (i)}\ \ \ \frac{1}{2}\widetilde{\bm w}B^{-1}\bm v,\notag \\
&{\rm (ii)}\ \ \  \frac{1}{2}\widetilde{\bm w}B^{-1}{\bm w}'
+\frac{1}{4}\widetilde{\bm w}B^{-1}\bm v\, \widetilde{{\bm w}'}B^{-1}\bm v ,\notag \\
&{\rm (iii)}\ \ \  \frac{3}{2}{\rm Tr}B^{-1}C+\frac{1}{4}\widetilde{\bm v}B^{-1}C B^{-1}\bm v,\notag \\
&{\rm (iv)}\ \ \  \frac{1}{2}\widetilde{\bm w}B^{-1}CB^{-1}\bm v+\frac{3}{4}(\widetilde{\bm w}B^{-1}\bm v) \, {\rm Tr}B^{-1}C \notag \\
&\qquad \ \ +\frac{1}{8}(\widetilde{\bm w}B^{-1}\bm v) (\widetilde{\bm v}B^{-1}CB^{-1}\bm v),
\notag \\
&{\rm (v)}\ \ \  \frac{3}{2}{\rm Tr}B^{-1}CB^{-1}C'+\frac{1}{2}\widetilde{\bm v}B^{-1}CB^{-1}C'B^{-1}\bm v \notag \\
&\qquad \ \ +\Big(\frac{3}{2}{\rm Tr}B^{-1}C+\frac{1}{4}\widetilde{\bm v}B^{-1}CB^{-1}\bm v\Big)\notag \\
&\qquad \ \ \times 
\Big(\frac{3}{2}{\rm Tr}B^{-1}C'+\frac{1}{4}\widetilde{\bm v}B^{-1}C'B^{-1}\bm v
\Big).
\label{list.P_O}
\end{align}

As an example, we show how to obtain 
the matrix element of $\widetilde{\bm x}C\bm x$ 
belonging to class (iii). The corresponding ${\cal P_{O}}$ reads 
\begin{align}
\frac{3}{2}{\rm Tr}B^{-1}C+\frac{1}{4}\widetilde{\bm v}G_1\bm v
\end{align}
with $G_1=B^{-1}CB^{-1}$, and it comprises four terms:
\begin{align}
&T_{000}=\frac{3}{2}{\rm Tr}B^{-1}C,\ \ \  T_{100}=\frac{1}{4}\widetilde{u}G_1u,\notag \\
&T_{010}=\frac{1}{4}\widetilde{u'}G_1u',\ \ \  T_{001}=\frac{1}{2}\widetilde{u}G_1u',
\end{align}
where $T_{kk'l}$ is an abbreviation of $T_{kk'l}(u'A', uA)$. 
Equation~(\ref{me.vs.ovlme}) 
immediately gives us the following result:
\begin{align}
& \langle f^{u'A'}_{K'LM} | \widetilde{\bm x}C\bm x | f^{uA}_{KLM} \rangle \notag \\
&= \langle f^{u'A'}_{K'LM} | f^{uA}_{KLM} \rangle 
\Big[\  \frac{3}{2}{\rm Tr}B^{-1}C+\frac{\widetilde{u}G_1u}{\widetilde{u}B^{-1}u}F^{KK'L}_{100}(z)\notag \\
&\ \  +\frac{\widetilde{u'}G_1u'}{\widetilde{u'}B^{-1}u'}F^{KK'L}_{010}(z)+\frac{\widetilde{u}G_1u'}{\widetilde{u}B^{-1}u'}F^{KK'L}_{001}(z)\Big].
\label{me.xcx}
\end{align}

The matrix element of $(\widetilde{\bm x}C\bm x)^2$ is obtained 
similarly: 
\begin{widetext}
\begin{align}
& \langle f^{u'A'}_{K'LM} | (\widetilde{\bm x}C\bm x)^2 | f^{uA}_{KLM} \rangle= \langle f^{u'A'}_{K'LM} | f^{uA}_{KLM} \rangle  \Big[\,  \frac{3}{2}{\rm Tr}G_1C+\frac{9}{4}({\rm Tr}B^{-1}C)^2  
+(2\widetilde{u}G_2u+3{\rm Tr}B^{-1}C\widetilde{u}G_1u)\frac{1}{\widetilde{u}B^{-1}u}F^{KK'L}_{100}(z)\notag \\
&+(2\widetilde{u'}G_2u'+3{\rm Tr}B^{-1}C\widetilde{u'}G_1u')\frac{1}{\widetilde{u'}B^{-1}u'}F^{KK'L}_{010}(z)
+(2\widetilde{u}G_2u'+3{\rm Tr}B^{-1}C\widetilde{u}G_1u')\frac{1}{\widetilde{u}B^{-1}u'}F^{KK'L}_{001}(z)\notag \\
&+\frac{(\widetilde{u}G_1u)^2}{(\widetilde{u}B^{-1}u)^2}F^{KK'L}_{200}(z)+
\frac{(\widetilde{u'}G_1u')^2}{(\widetilde{u'}B^{-1}u')^2}F^{KK'L}_{020}(z)
+ \frac{(\widetilde{u}G_1u')^2}{(\widetilde{u}B^{-1}u')^2}F^{KK'L}_{002}(z)\notag \\
&+2\frac{(\widetilde{u}G_1u)(\widetilde{u'}G_1u')}{(\widetilde{u}B^{-1}u)(\widetilde{u'}B^{-1}u')}F^{KK'L}_{110}(z)
 + 2\frac{(\widetilde{u}G_1u)(\widetilde{u}G_1u')}{(\widetilde{u}B^{-1}u)(\widetilde{u}B^{-1}u')}F^{KK'L}_{101}(z)
+ 2\frac{(\widetilde{u'}G_1u')(\widetilde{u'}G_1u)}{(\widetilde{u'}B^{-1}u')(\widetilde{u'}B^{-1}u)}F^{KK'L}_{011}(z)
\Big],
\label{me.xcx**2}
\end{align}
\end{widetext}
where $G_2=B^{-1}CB^{-1}CB^{-1}$.

\end{document}